\RequirePackage{fix-cm}
\documentclass[]{svjour3}       
\smartqed  
\usepackage{natbib}
\usepackage{booktabs} 
\usepackage{multirow}
\usepackage{makecell}
\usepackage[ruled]{algorithm2e} 
\usepackage{color,soul}
\usepackage[colorinlistoftodos]{todonotes}
\usepackage{paralist}
\usepackage{caption}
\usepackage{hyperref}
\usepackage{subcaption}
\usepackage{graphicx}
\usepackage[utf8]{inputenc}
\usepackage{tikz, times}
\usepackage[final]{pdfpages}
\usepackage{textcomp}
\usepackage{multicol}

\usepackage[colorinlistoftodos]{todonotes}

\makeatletter
\newcommand\footnoteref[1]{\protected@xdef\@thefnmark{\ref{#1}}\@footnotemark}
\makeatother

\usetikzlibrary{decorations.pathreplacing}
\definecolor{myLightGray}{RGB}{191,191,191}
\definecolor{myGray}{RGB}{160,160,160}
\definecolor{myDarkGray}{RGB}{144,144,144}
\definecolor{myDarkRed}{RGB}{167,114,115}
\definecolor{myRed}{RGB}{255,58,70}
\definecolor{myGreen}{RGB}{0,255,71}
\usetikzlibrary{mindmap,fit,backgrounds,shadows,arrows,positioning}
\tikzset{
    >=stealth',
    punkt/.style={
           rectangle,
           rounded corners,
           draw=black, very thick,
           text width=6.5em,
           minimum height=2em,
           text centered},
    pil/.style={
           ->,
           thick,
           shorten <=2pt,
           shorten >=2pt,}
}

\tikzset{grow cyclic list/.code={%
  \def\tikzgrowthpositions{{#1}}%
  \foreach \n [count=\i,remember=\i]in {#1}{}%
  \let\tikzgrowthpositionscount=\i%
  \tikzset{growth function=\tikzgrowcycliclist}}}
\def\tikzgrowcycliclist{%
  \pgftransformshift{%
    \pgfpointpolar{\tikzgrowthpositions[mod(\the\tikznumberofcurrentchild-1,\tikzgrowthpositionscount)]}%
      {\the\tikzleveldistance}}}

\begin{document}

\title{Experiential Learning Approach for Software Engineering Courses at Higher Education Level}



\author{Javier Gonzalez-Huerta\and         
        Jefferson~Seide~Moll\'eri\and
        Aivars \v{S}\={a}blis\and
        Ehsan Zabardast}


\authorrunning{Gonzalez-Huerta et al.} 

\institute{J. Gonzalez-Huerta, A. \v{S}\={a}blis {and}  E. Zabardast   \at
              Software Engineering Research Lab Sweden\\ 
              Blekinge Tekniska H\"{o}gskola\\
              Valhallav\"{a}gen 1\\ 371 79,  Karlskrona, Sweden\\
              \email{\{javier.gonzalez.huerta, aivars.sablis, ehsan.zabardast\}@bth.se}             \\
             \emph{Corresponding  address:} of J. Gonzalez Huerta 
          \and
          Jefferson~Seide~Moll\'eri\at
              Simula Metropolitan Centre for Digital Engineering\\Oslo, Norway\\ 
              \email{jefferson@simula.no}
}

\date{Received: date / Accepted: date}

\maketitle

\begin{abstract}

\textbf{Background:} Software project management activities help to introducing software process models in Software Engineering courses. However, these activities should be adequately aligned with the learning outcomes and support student's progression. 

\textbf{Objective:} Present and evaluate an approach to help students acquire theoretical and practical knowledge and experience real-world software projects' challenges. The approach combines a serious game and a design-implement task in which students develop a controlled-scale software system.

\textbf{Method:} To evaluate our approach, we analyzed the students' perceptions collected through an online survey, their project plans, and their final reports using thematic analysis.

\textbf{Results:} Results suggest that the approach promotes knowledge acquisition, enables students' progression, reinforces theoretical concepts, and is properly aligned with the course's learning outcomes.

\textbf{Conclusion:} The approach seems to help introducing software process models in Software Engineering courses. Our experience can also be inspiring for educators willing to apply our approach in similar courses.
\end{abstract}

\section{Introduction}

Introducing students to Software Engineering (SE) is a challenging task~\citep{Broman2010,Malik2012}, that needs to integrate more than just classroom-based teaching and learning activities~\citep{Malik2012,Yadav2010}.

Project management activities are intended to provide students with a dynamic environment that mirror real-world challenges~\citep{bruegge1991software}. This problem-based learning approach is supported by the pedagogical theories for knowledge acquisition through ``learn by doing''~\citep{dewey1897my}, and experiential learning~\citep{kolb2014experiential} philosophies, as well as a student-focused approach to teaching~\citep{hung2011theory,flener2006realism}.

A major issue associated to project management courses or to \textit{design-implement} tasks in the software engineering curriculum is how to find a balance between the level of realism, and the relevance of the contents students will learn \citep{flener2006realism, bruegge1991software}. On the one hand, toy projects failed to connect theory taught in the classroom to real-world problems. On the other hand, real-world project conditions are often too demanding to fit a course schedule. Simulations have been used to achieve the balance without incurring these issues \citep{peterson2011teaching}.  

From our experiences with \textit{design-implement} tasks in Software Engineering courses, we would also add another risk: that students are more focused on the technical challenges of the project tasks \citep{bruegge1991software}. Thus, they are sometimes willing to ``hack'' the solution instead of focusing on the software development practices and models they are intended to be following. This focus on hacking the solution might prevent them from experiencing and reflecting on the intended learning outcomes (ILOs).

Therefore, at Blekinge Institute of Technology (BTH), we are currently developing an integrated approach using a game activity and a development task to expose students to a software development project. The approach simulates some of the challenges faced when planning and managing real-life software project )but in a controlled scale).

We implemented and empirically evaluated our approach in the context of a software engineering course for two consecutive years. This paper presents the results of this evaluation, in which we analyzed the adherence of their project plans to a set of rubrics that drive the serious game, analyzed the teams’ reports submitted at the end of the course, as well as students' perceptions regarding the integrated learning approach, gathered through an online survey.

The paper is structured as follows: Section \ref{sec:background} presents the related work. Section \ref{sec:methods} describes the research method. Section \ref{sec:results} provides the results, which are further discussed Section \ref{sec:discussion}. Finally, Section \ref{sec:conclusions} concludes the paper.

\section{Background \& Related Work}\label{sec:background}

In this section, we introduce some of the techniques used for teaching Software Engineering and Software Project Management, the use of games in Software Engineering education, and the use of Design-Implementation tasks in Software Engineering courses.

\subsection{Teaching Software Engineering and Software Project Management}


Software Engineering is defined in ISO/IEC/IEEE 24765:2010~\citep{ISO24765} as \textit{``The application of a systematic, disciplined, quantifiable approach to the development, operation and maintenance of software; that is, the application of engineering to software.''}. The ACM/IEEE Curriculum~\citep{ACM2014} provides guidelines for the definition of Software Engineering education at undergraduate level, and identifies the main expected student outcomes, the Software Engineering Education Knowledge (SEEK). The guidelines are explicit when recommending that \textit{``The curriculum should have significant real-world basis''}~\citep{ACM2014}, and suggest to incorporate at least some real-world related activities they identify, such as case studies, project-based activities, practical exercises, student work experiences or capstone projects.

However, in any case, teaching Software Engineering is a challenging task \citep{Malik2012,Broman2010}. Among other challenges, different authors identify the fact that only classroom-based teaching and learning activities are not enough, and might be an ineffective approach towards SE teaching \citep{Malik2012,Yadav2010}.

If we look at research on Software Engineering Education, several Systematic Mapping Studies, e.g., the ones presented \cite{Marques2015,Malik2012}, identify Problem-Based Learning as crucial activities in Software Engineering Education. Problem-Based Learning (PBL) promotes active learning and knowledge acquisition through group work \citep{hung2011theory, schmidt1994problem}. PBL is a teaching practice related to the constructivist theories, in which the student is directly responsible for the knowledge construction \citep{hendry1999constructivism,elmgren2014academic}, a.k.a. student-focused approach. It aims to produce knowledge by connecting the students' prior knowledge to new facts and understanding. PBL extensively uses reflection, critical thinking, and experimentation as learning facilitators. In this practice, students are presented with a situation that requires a solution, whereas the teachers act as supervisors (and sometimes simulated customers), stirring the group toward a potential solution. The PBL takes several meetings, and in the time between meetings, students should look to deepen their knowledge regarding the problem.

\begin{figure}[!ht]
\centering
\caption{Learning cycle by \cite{kolb2014experiential}.}
\includegraphics[width=0.70\textwidth]{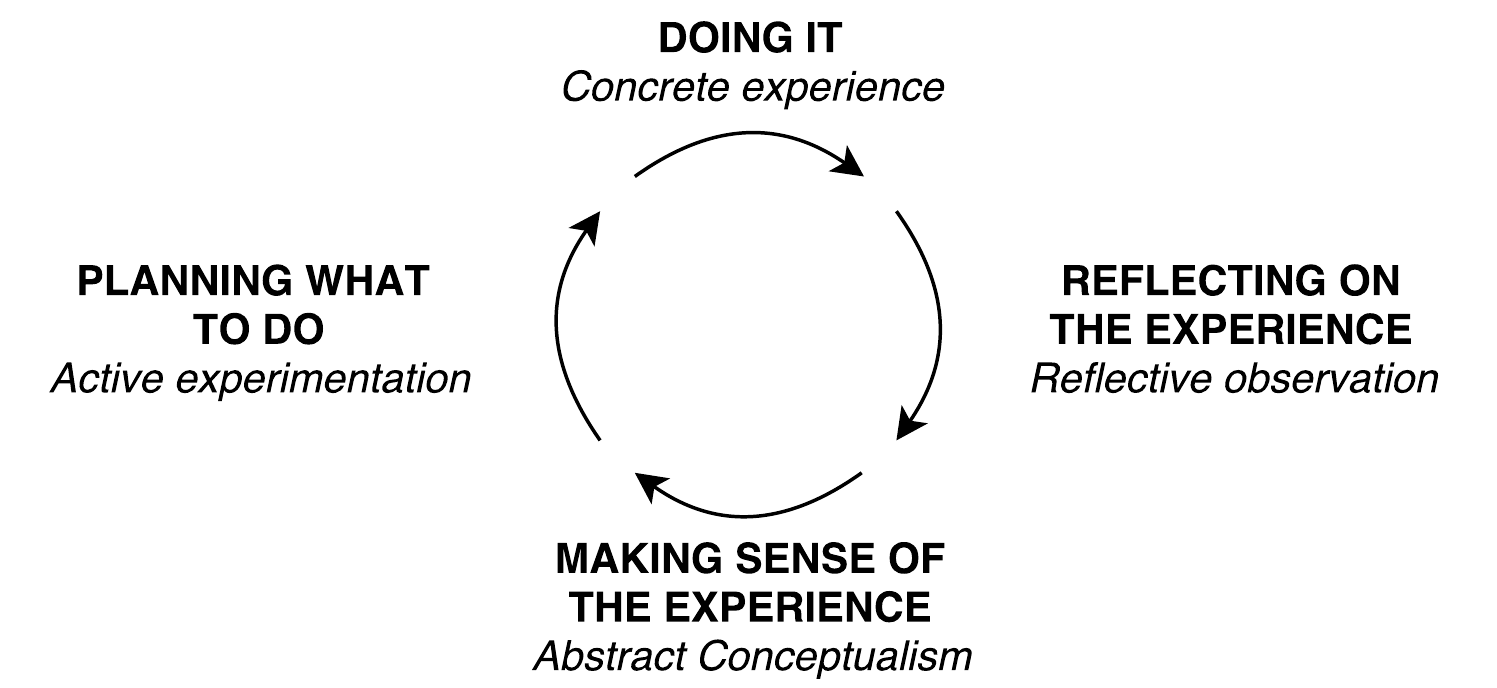}
\label{fig:learningcyclekolb}
\end{figure}

\cite{kolb2014experiential} describes those stages in a experiential learning cycle (illustrated in Figure~\ref{fig:learningcyclekolb}). In the first stage, i.e., doing it, the student is faced with a new experience, herein the game challenge. In a group, the student is foster to reflect on the challenge (stage 2) and make sense on a candidate solution (stage 3). Finally, the student applies the solution and gather its results. The cycle starts again, as the student progresses into a deeper understanding of the topic.

Both PBL and Experiential Learning are connected to the ``learn by doing'' philosophy introduced by John~\cite{dewey1897my}. These views favor relevant and practical learning and are particularly useful for skill-oriented engineering courses/programs. 

One suitable solution to enable this ``learn by doing'' philosophy in Software Engineering education is to have dedicated Software Project Management Courses (PMCs) to expose the students to the specificities of Software Project Management. PMCs have become increasingly popular in Software Engineering education, particularly at the end of undergraduate and subsequent graduate programs \citep{ralph2018re, broman2012company}. Ideally, PMCs should be aligned to PBL and related pedagogical philosophies but this is not always the case~\citep{ralph2018re, flener2006realism}. When students get exposed to Problem-Based Learning in Software Engineering education in basic PMCs, the problems tend to be adapted and simplified, and often linked to ``pre-fabricated'' solutions~\citep{Oliveira2013,Souza2018}, or extensive real-client projects that are likely to overload students with technical issues~\citep{koolmanojwong2009using, bruegge1991software}, and students perceive that they are not exposed to ``real'' project management~\citep{Kruchten2011}. 

Besides, our experiences using PBL, through design and implement tasks in Software Engineering introductory courses, is that the students tend to focus on the technical problem itself, trying to ``hack'' the solution, rather on the underlying process and methodology, that is the main focus of such courses. In software projects formal and familiar technical aspects can be against with human and sociological factors~\citep{Caulfield2011}. Monitoring the extent to which students follow the processes and methods is challenging, and might require additional monitoring by instructors, e-learning portfolios, or additional reporting activities by the students~\citep{Marques2018}. This is why simulations and serious games might help the students to experience some challenges of Software Project Management, without the complexity of developing the solution. 

\subsection{Games in Software Engineering Education}

Serious Games and Simulations have been lately become popular as tools for experiencing some of the challenges of software development projects~\citep{Marques2015,Souza2018,Kosa2016}, and the Game-Based Learning~\citep{Wangenheim7} has been even defined in the context of Software Engineering Education. Recently, its usage has received bigger attention, and has been even the focus of Systematic Literature Reviews and Systematic Mapping Studies, e.g., in~\citep{Kosa2016,Souza2018,Marques2015}.

The term “Game Based Learning” (GBL)~\citep{Wangenheim7} refers to any approach using games for learning purposes. \cite{Wangenheim7} define the terms as the use of game applications for defining learning outcomes. Games are any contest (play) among adversaries (players) operating under constraints (rules) for an objective (winning, victory, or pay-off) \citep{Wangenheim7}. While in competitive games players are opposed to each other, cooperative games encourage players to work together for mutual benefit. Collaborative games go a step further, and participants work together as a team towards a common goal~\citep{zagal2006collaborative}.

One of the rationales behind the relatively wide usage of Simulated or Game Based Learning is the complex nature of the topic and the time restrictions imposed by the schedule of the courses~\citep{Souza2018}.  

Serious Games have been used to illustrate software project management~\citep{raabe2013serious, petri2017quality}, software development processes~\citep{hainey2011evaluation,navarro2004simse,benitti2008utilizaccao,baker2003problems,baker2005experimental,Souza2017,Souza2018}, to teach Risk Management in Software Engineering Projects e.g.,~\citep{Taran2007,Oliveira2013}, to introduce the usage of Kanban or Scrum in Software Engineering Projects e.g.,~\citep{Heikkila2016,Paasivaara2014,Fernandes2010}, to introduce Requirements Engineering, e.g.,~\citep{Knauss2008,hainey2011evaluation} or to illustrate the particularities of Global Software Engineering e.g.,~\citep{VanSolingen2011,Sablis2019}.

However, the majority of theses serious games are intended to be played in a single class-session. Thus, this setup is not entirely aligned to the PBL approach, that requires the students: 
	\begin{inparaenum}[i)]
		\item to search for candidate solutions to a problem, 
		\item to critically analyze the candidates in order to make a decision,
		\item and finally to reflect on the impacts of such potential solution.
	\end{inparaenum}
Which are indeed required skills to be trained in Engineering education~\citep{Crawley2014}. Moreover, the need to integrate all the game in a single session tends to oversimplify the tasks that can be carried out and the level of involvement by the students.

As an alternative, in our previous work, we have prototyped and piloted a Legacy Game in the context of a SE-PMC course~\citep{molleri2018legacy}. A Legacy Game is designed to dynamically change throughout of a series of sessions~\citep{daviau2017legacy}. New game rules and contents can be introduced during the execution of the game, and similarly, old content may get overridden or removed~\citep{bbg2018glossary}. Our legacy game incorporates the concepts of PBL into a project management course. It is designed to illustrate two different development models (i.e., plan-driven and agile).

However, some authors argue that, although games are useful pedagogical tools and well received by students, they might not be enough, and other learning activities might be needed to reinforce the learning~\citep{Caulfield2011,Souza2018,Heikkila2016}. This is the reason why we have defined an integrated approach that combines traditional lectures, serious games, as well as design-implement task.  In the design-implement task, the students can experience some of the concepts and challenges introduced in the serious game. These concepts and challenges are later framed in the final report in which students discuss their experiences in the course, which has the goal of increasing the effects on student's learning, as suggested by~\cite{Hult2000} and~\cite{Enstrom2014}. 

\subsection{Design-Implement Tasks}

Providing students with real-world experiences in an academic setting might help students to understand problems they will face once they go to the industry, as well as experience challenges when working in software development teams and difficulties and complexities when working in real-world software development environment~\citep{bavota2012teaching}. However, in an academic setting, it is challenging to provide these experiences, where students can deal with issues that arise from common projects, such as realistic requirement management, coping with time pressure and dealing with software and hardware constraints~\citep{Marques2018}.  

Typically PMCs in software development in academic settings range from weeks-long assignments to full-semester activities and from solving problems that students free-to-choice to working with companies with problems that companies did choose to assign to students. However, these approaches typically allow simulating only part of real-world experiences. Providing students with real-world experiences involving realistic requirement management, working with a customer, and dealing with deadlines is a complex challenge. Some of these projects are hard to find (i.e., with full scope and full end-to-end process for just a team with limited time) in real-world settings.  And finally, PMCs in academic settings that offer more realistic environments tend to become complicated and require a lot of coordination~\citep{johns2013simulating}. In our programs, more realistic settings are achieved through capstone projects in big teams working on industrial projects, but this setup is not reasonable in introductory courses.

Another feasible approach is to design courses with \textit{design-implement} tasks, following the terminology of the CDIO (Conceive Design Implement Operate) Framework~\citep{Crawley2014}. Design-implement experiences are, in the CDIO context, a ``range of engineering activities central to the process of developing new products or systems''~\citep{Crawley2014}. CDIO's design-implement experiences aim at providing students with practical experiences that consolidate theoretical knowledge and support their learning about the engineering process. The experiences are implemented by means of teaching-learning activities (TLAs) which allow the students to go through the planning and implementation stages of a project~\citep{westphal2018course}. Such experiences can also integrate disciplinary knowledge with development of soft skills such as personal skills, social skills, and communication skills~\citep{Crawley2014, westphal2018course}. These real-world experiences also offer means for teachers to monitor student progress and provide a development environment with realistic and complex technical challenges, yet scalable in terms of student skills and knowledge in software development.

Thus, we have embedded a design-implement task in our proposed approach, that uses LEGO Mindstorms\textsuperscript{TM} to simulate a real-world environment regarding technical challenges in software development. LEGO Mindstorms\textsuperscript{TM} has been used previously in higher education to teach students in advanced software engineering project courses~\citep{Lew2010,Weissberger2014}, or embedded systems~\citep{kim2009introduction}. 

Our approach aims at providing students with a \textit{design-implement} task that resembles real-world project case, without part of its complexity. This \textit{design-implement} task is designed to illustrate and experience software project management in simulated real-world setting following agile software development practices.

\section{An Integrated Experiential Learning Approach for Software Engineering Courses}
\label{sec:approach}

In previous instances of our project management component in a Software Engineering course, the students worked in a fully-fledged, traditional design-implement task as the sole \textit{``experiential''} teaching-learning activity. This task was reported by students in course retrospectives and course evaluations as too technical and too difficult, as most of them lack deep development experience. Thus, students sometimes did not achieve the intended learning outcomes through real design-implement tasks, as they were too focused on solving the technical challenges.

Therefore, our main objective is to propose an integrated experiential learning approach for SE courses that first simulates a real-life software development process's challenges. After that, the integrated approach comprises a design-implement task that allows students to experience these challenges we simulated, but this time in a real, still small-scale design-implement task. Students are expected to handle requirements elicitation, project planning, effort estimation, development, and testing while managing unexpected situations and uncertainty. From a pedagogical perspective, we expect the students to demonstrate the knowledge acquired with this experiential learning approach by the end of the course.

The remaining of the section is structured as follows: in Sub-section~\ref{context} we describe the details of the course before implementing the intervention; in Sub-section~\ref{approach} we present the experiential learning approach; and finally, in Sub-section~\ref{execution} we present the different executions of the approach.

\subsection{Context: The Course Before the Intervention}\label{context}

The experiential learning approach has been integrated as part of a Software Engineering course, comprising of 7.5~ECTS\footnote{European Credit Transfer System}. The course's overall goal is to give the student basic knowledge of software engineering, the software development processes, and their main phases. The course contents also include the different software development models and practices and their impact at the product, process, and organizational levels.

The course provides both theoretical knowledge and its application in practical situations. Theoretical knowledge is provided in a series of lectures covering themes such as requirements elicitation and management, testing, architecture design, project planning, and project follow-up. The practical component requires the student to participate in the planning and execution of a small project.

This course was an interesting opportunity for our integrated experiential learning approach that combines a game activity \citep{molleri2018legacy} with a real design-implement task of a certain but controlled size. Although the SE course shares certain similarities with software project management courses, it is a core-area course that uses a small-scale, controlled-environment TLA for the students to experience a software project's challenges.

\begin{figure}[!ht]
\centering
\caption{Visual representation of the course components according to constructive alignment. ILOs are connected to the related TLAs and ATs via solid, dashed and dotted line, each line corresponding to a ILO.}
\includegraphics[width=0.8\textwidth]{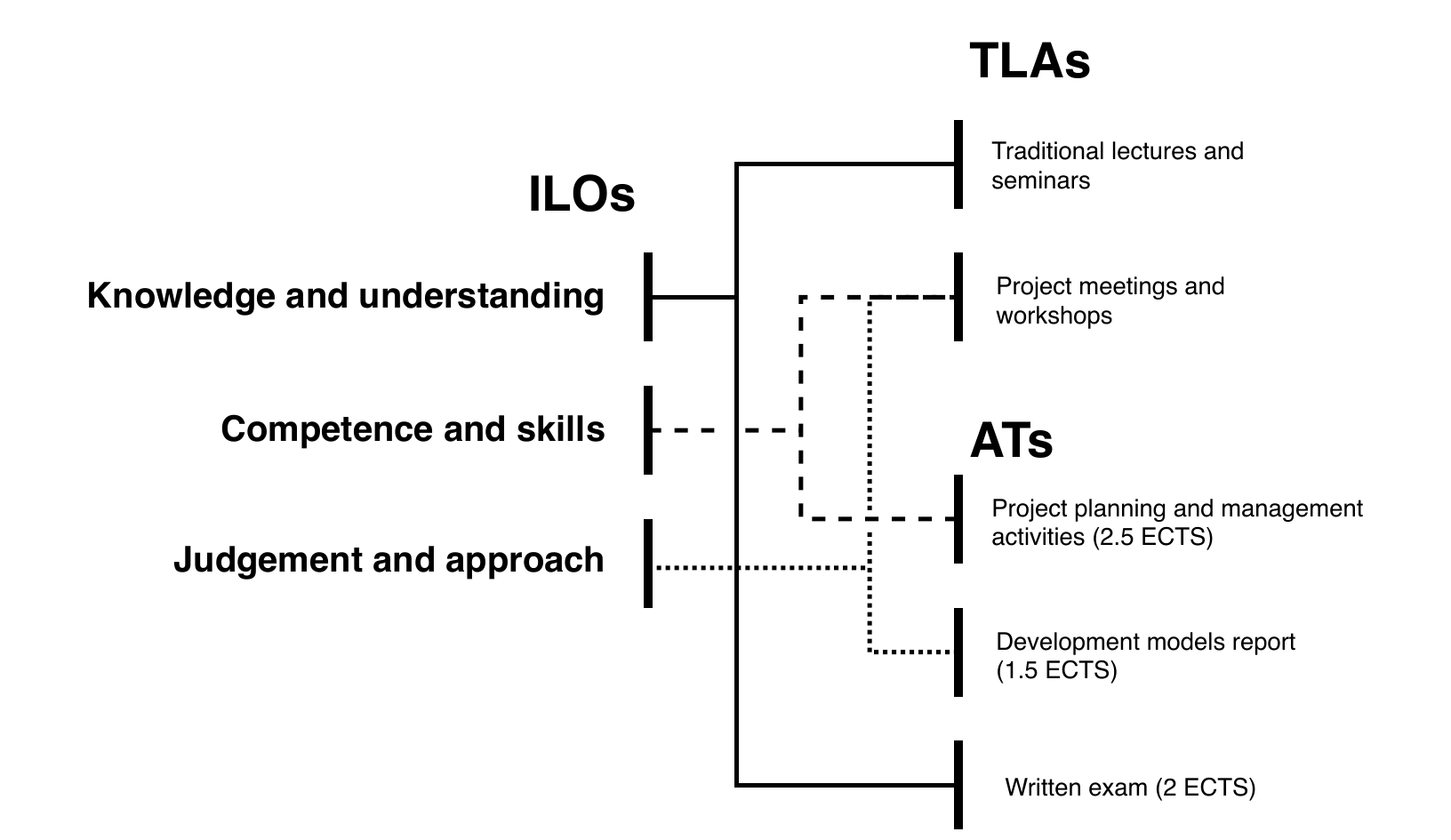}
\label{fig:constructiveAlignment}
\end{figure}

Figure~\ref{fig:constructiveAlignment} illustrates the main components of the course comprising Intended Learning Objectives (ILOs), Teaching Learning Activities (TLAs), and Assessment Tasks (ATs), according to the constructive alignment framework~\citep{biggs2014constructive}.

\subsubsection{Intended Learning Objectives}
\label{sec:ilos}

The course goals are twofold. On the one hand, there is a focus on theoretical knowledge acquisition; and on the other hand, on its application in practical, small-scale, controlled examples. The course objectives, as described in the course syllabus, are:

\begin{itemize}
    \item \textbf{Knowledge and understanding.} The course aims to provide the student with knowledge about how software systems are developed. On completion of the course, the students will gain an understanding with regards to the main phases of software-intensive product development, the different processes, their strengths and weaknesses, practices, and methodologies in software development.
    \item \textbf{Skills and abilities.} The course provides knowledge about the development processes, requirements management, testing, architectural design, project planning, and project management, equipping the student with basic knowledge for participating in the planning and follow up of a project in practice, according to the selected development process.
    \item \textbf{Judgment and approach.} The students should be able to discuss the pros and cons of each process and method, but also how their potential impact on the software product, its users, and the development organization. The students are also expected to be able to present and argue for ethical aspects regarding current trends and products in society.
\end{itemize}

\subsubsection{Teaching-Learning Activities}
\label{sec:tlas}

The course is organized around traditional lectures, seminars, debates, and group assignments. Students are expected to work individually as well as in teams, especially in the seminars and workshops.

\begin{itemize}
    \item \textbf{Theoretical lectures and seminars.} The lectures introduce the Software Engineering disciplines, different stages of software-intensive product development, development processes, the historical perspective and limitations of each of them. Besides, ethical aspects are also introduced and discussed in the context of software-intensive product development. The primary language of instruction Swedish, but seminars, some lectures, material, and reports can take place in English as well.
    \item \textbf{Project meetings and workshops.} The project planning and management activities, as well as the software development project require interaction between the teachers and teams of students. This interaction is organized around project meetings and workshops, in which the teams will discuss the project planning, management, and execution activities with regards to the software product being developed.
\end{itemize}

\subsubsection{Assessment Tasks}\label{assesment_task}
\label{sec:ats}

Both individual and group assignments are included in the course. The examination of the course was graded based on three different tasks: 

\begin{itemize}
    \item \textbf{Project planning, management and execution activities (2.5 ECTS).} Software project activities, in which the students, organized in teams of 3-4 participants, have to elicitate, document, and prioritize the requirements of a software system, document its design and plan and manage its execution, including a design-implement task. All activities are graded altogether as pass/fail and graded as a unique activity.
    \item \textbf{Development models report (1.5 ECTS).} At the end of the project planning, management, and execution activities, the students submit a summative reflective report summarizing the experiences with both the traditional and the agile models, pros and cons, when to use each one, and personal experiences. The report should include a discussion about ethical and professional practices in software development. This report is handled, submitted, and graded as a group (A-F).
    \item \textbf{Written exam (2 ECTS).} An individual on-line exam to sum up all the contents of the course, with the goal of having a way to assess the individual learning of each student. Although we hope students' performance been positively impacted by the integrated experiential learning approach, the written exam and its outcomes were not part of our study.
\end{itemize}

\subsection{Integrated Experiential Learning Approach}\label{approach}

\begin{figure}[!ht]
\centering
\caption{Course schedule and evaluation timeline. The top part describe the specific course units (i.e. TLAs, ATs), while the bottom part relate the data collection instruments of our empirical validation.}
\includegraphics[width=1\textwidth,trim={0 7.5cm 0 0},clip]{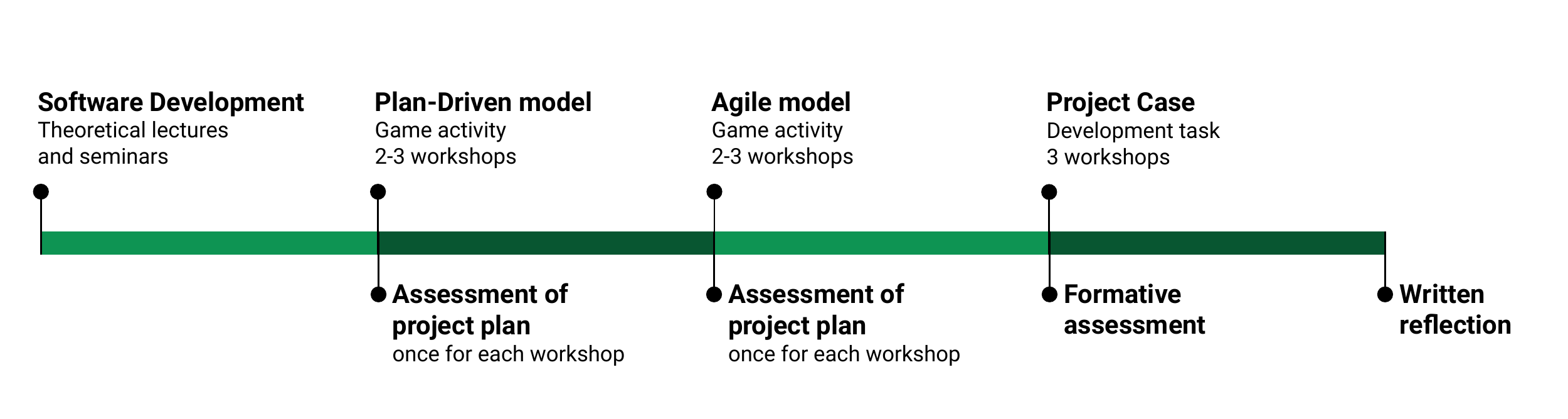}
\label{fig:datacollection}
\end{figure}

The proposed learning approach integrates two instances of a simulated project planning activity and one software design-implement task (see Project planning, management, and execution activities in Section~\ref{assesment_task}) as illustrated in Figure~\ref{fig:datacollection}. 

For the simulated project planning activity, we used our serious game~\citep{molleri2018legacy} twice, one in which the students follow a plan-driven development process, and one in which the students have to follow a self-tailored agile development process.  We run each simulated project planning activity once the theoretical contents had been covered. In this way, the game challenges match the knowledge the students were exposed to.

The first instance of the game activity focused on a traditional, plan-driven development process. The students worked on a project plan following a more formal plan-driven (a.k.a. waterfall) development process model. This traditional planning incorporates more extensive documentation, detailed plans, and risk management that leads to the process monitoring.

In the second instance of the game, students planned another project employing a self-tailored agile process model. The agile project incorporates strategies such as team interaction, customer collaboration, and response to changes. In this case, the students also need to find arguments for choosing the different agile practices and agile project management techniques they opted for.

After the two instances of the game activity, the students are exposed to actual software development, i.e., the design-implement task. This design-implement task involves students working with a controlled-scaled, real-world-like project case using an agile approach. The project case is in the same domain as students planned in the second instance of the game (i.e., the one following the agile process). In this way, the design-implement task matches both the theoretical knowledge the students have gained and the domain knowledge from the game activity's simulated project planning.

The actual performance on the simulated projects (i.e., whether they deviated from the original planning, or whether the planning was accurate) and on the design-implement task (i.e., the adherence to the requirements and the quality of the developed solution) is not part of the grading of the course. The main reason is that the main learning outcome is not about the programming skills of the students, but their ability to plan and follow a development process. The pass or fail is based on their participation on the different activities (i.e., project meetings and workshops), and the degree to which the students where following the development process selected, which was judged and assessed by the teaching staff.

\subsubsection{Game Activity}\label{gamerules}

Students worked in teams of 3-4 participants that shared a common goal during the games activities. Our serious game can be classified as collaborative according to~\citep{zagal2006collaborative} classification, since all participants pursue the same goal (i.e., there is no competition or competing goals). The serious game comprises a previously-defined number of turns based on the course schedule. We carried out a workshop with students as a kick-off session at the beginning of the game activity. In the kick-off session, teachers presented the project description and set up of the game and the number of turns the game will comprise. At the end of the kick-off session, there was a requirements elicitation session, in which each designated member of the teaching staff acted as a customer/product owner for each team. The different teams were able to ask for clarifications.

Each turn, which corresponds to one- to two-week cycles, includes five steps. The first two steps took place outside the classroom, starting right after the kick-off session. The remaining steps were carried during the next workshop session at the end of a game turn. The game steps were as follows~\citep{molleri2018legacy}:

\begin{enumerate}[1)]
  \setlength{\itemsep}{1pt}
  \setlength{\parskip}{0pt}
  \setlength{\parsep}{0pt}
  \item At the beginning of each game turn, each team worked on and submitted a \textbf{project plan}. The project plan should meet the project description described by the teacher, also detailing resources (e.g. budget, workforce) and constrains (e.g. time to deliver, business rules).
  \item Then, the teachers \textbf{assessed the submissions} with regards to the intended learning objectives. If the team's project plan was sounding, adhered to good practices, and matched the project's needs, the team was awarded bonuses for the current game turn. On the contrary, the team was penalized if their project plan omitted certain required development activities or practices or included well-known bad practices. We have designed a set of rubrics to assess both bonuses and penalties\footnote{The rubrics are available as an appendix to the game rules in \url{https://goo.gl/1oUvvB}}.
  \item During the workshop session, the game mainly consisted of \textbf{rolling dices that represented the uncertainties} of a software development project. Teachers guided the students during the whole step, relating these uncertainties to real examples, e.g., a penalty in the implementation could be caused by a non-updated design or lack of requirements traceability. The uncertainty values were then added to the bonuses/penalties scored by the team's project (see step 2 above). This final result represents how much the actual process deviates from the original plan.
  \item Further, teachers \textbf{presented project changes and new challenges} to the teams, representing events that are likely to occur in a software process, such as new requirements or resource limitations. Teachers played the role of customers or product owners, fostering students to negotiate the inclusion of some of the new requirements or resource allocation.
  \item Finally, the teams were asked to \textbf{update the project plans} according to the deviation and new events resulting from steps 3 and 4. They were also suggested to discuss the challenges and make any improvements they thought were needed for the project's success. As an example, one team reflected on their updated plan as follows: \textit{``The decision we had to take was, deliver the minimum viable product in 5 sprints or be late. We chose to push for the minimum viable product and did not deal with technical debt or the bugs''}.
\end{enumerate}

In each game workshop session, the teaching staff provided feedback to students, expecting them to reflect on their choices and learn from the mistakes. At the end of each game,  the teams were asked to provide their reflections regarding the process, the outcome, and perceptions about the gaming experience and lessons learned. 

\subsubsection{Design-Implement Task}

The design-implement task started after the end of the Serious Game. The goal of this task was to develop the same software system as planned in the agile instance of the serious game. Students were expected to reuse parts of their project plans when carrying out the design-implement task. This set up allowed students to start the design-implement task with certain domain knowledge and to experience how different it is to only plan something compared to developing the solution themselves.

The students had to develop different software components that together integrated a self-driving system for LEGO Mindstorms\textsuperscript{TM} vehicles (trucks, cranes, forklifts) controlled through EV3 Blocks using ev3dev/JAVA. The software components responded to different requirements expressed through User Stories, with pre-and post-conditions. The range of complexity ranged from more simple features like controlling the motor's speed to more complex stories like \textit{drive-following-a-line} or epic stories.  One example of an epic story was self-driving the vehicle until reaching a specific location where they had to interact with another vehicle. In the \textit{drive-following-a-line} story, teams had to develop routines to detect a line in the floor and drive the vehicle following that line, keeping the line centered in the vehicle until the sensors read a stop signal (i.e., a transversal line filled with different color/pattern).

The design-implement task was planned to comprise three Scrum development sprints, with a two-week sprint lead time. As part of the design-implement task, four workshops were conducted with students: the first as a kick-off session and three after each sprint. In the kick-off session, the teacher presented the design-implement task description and the task's goal to the teams. In this session, the teaching staff presented the tools to be used, the code provided as scaffolding, and the technical details needed to carry out the development work and constraints (e.g. hardware, software). The teaching staff also acted as Customers/Product Owners negotiating with each team the Minimal Viable Product (MVP) and prioritized the user stories.

In the remaining three workshops, the teams discussed and planned releases. Teams did estimation based on story points and user story prioritization together with teachers acting as product owners. The second and third workshops were organized as a sprint review. In this sprint review, each team showed the progress done stories implemented during the sprint, adapted the backlog, calculated the velocity,  and planned the next sprint. The fourth and last workshop was organized as a release demo, in which the students did a live demo of the user stories they have implemented, with the teaching staff acting as customers and product owners.

During the design-implement task, the students were able to book time in the lab to provide them with access to hardware and supervision and mentoring by the teaching staff. Teams could book access to the lab for 2 hours per week upon registration, although no strict policy enforcing the access to only two hours was put in place (only when conflicts arose on specific dates). Students were encouraged to work off-line and use the time in the lab to test their solutions. Every team had their Scrum boards visible in the lab so that the teaching staff could monitor the progress of the different teams.

\subsection{Executing the Integrated Experiential Approach}\label{execution}

We executed the experiential approach in a Software Engineering course at 
BTH for two consecutive years, i.e., 2018 and 2019. Each year, we have conducted two instances of the game activity for two different software development processes and one instance of the design-implement task, as illustrated in Figure~\ref{fig:datacollection}. The assignments and rubrics were slightly different in each case.

We informed the students beforehand that the game's outcome was not part of the course evaluations. So, the students were not penalized in their grades if their project did not get particular bonuses. One could expect that some students use that as an excuse for not working hard. However, we soon realized that the gamification of the exercise produced a motivational effect. In general, students were trying their best to attain the best possible results, sometimes showing frustration when they were unlucky with the dice-rolls.

\subsubsection{2018}

Our integrated experiential learning approach was first put in place during Spring 2018. Fifty-eight students divided into fourteen teams took part in the game and the design-implement task. They were registered in a five-year  Master of Science in Engineering program (300 ECTS) in Industrial Economy\footnote{Civilingenjor i Industriell Ekonomi}. Forty students were in their second year,  while eighteen students were in their third year pursuing a specialization in Software Engineering and Information Technology. Both for the students in their second and third year, the course becomes their introduction to Software Engineering. The second-year students have read their first course in programming. In contrast, the third-year students had read the introductory course to programming in Java, a course in object-oriented programming,  and a course in Database Engineering. 

This mixture was due to a change in the program's structure. The course was moved from the third to the second year, and 2018 was the year we met both cohorts at the same time. 

There were no mixed groups, i.e., there were no groups with students in their second working together with students in their third year. Although initially, we thought having mixed teams might be a good idea, in practice, that meant that students did not have overlapping free slots in their schedules to work together, and therefore they chose to have homogeneous teams. 

Both the game activity and the design-implement task were supervised by at least two persons in each session. During the game activity, we further divided the 14 teams into two different groups with seven teams each. In other words, We conducted all workshops twice, for each group of seven teams. This strategy allowed the teaching staff to provide closer support for students without overloading them. The Lab time was supervised by at least one member of the teaching staff, although often two members were present providing support to students.

\subsubsection{2019}

The second application of the integrated approach took part in Spring 2019. Twenty-four students were initially divided into eight teams and participated in the integrated approach. After the plan-driven workshops, one of the teams dissolved, and we continued with seven teams. All students were registered in a five-year Master of Science in Engineering program (300 ECTS) in Industrial Economy. All students were in their second year, where the course became their introduction to Software Engineering after reading their first course in programming. In this execution, we managed all the groups simultaneously. However, we ran the workshops in two adjoining lab rooms since the serious game activities became too noisy, as some of the students pointed out in their feedback. The students again were provided with access to the lab, where teaching staff supported and mentored them when they experience technical difficulties.

It is important to mention that a change was introduced in the teaching staff, with one of the members taking a more leading role, since one of the team members was not part of the staff in 2019.

\section{Empirical Evaluation of the Approach}\label{sec:methods}

We have also conducted an empirical study to gain evidence regarding the extent to which our integrated approach is an effective means to enable students learning concerning the course objectives and contents.

\subsection{Research Objectives}

Our empirical evaluation aims at providing answers to the following research questions:

\begin{enumerate}
  \setlength{\itemsep}{1pt}
  \setlength{\parskip}{0pt}
  \setlength{\parsep}{0pt}
  \item[\textbf{RQ1}] How do students perceive that the integrated approach matches the learning objectives of the course?
  \item[\textbf{RQ2}] How does the integrated approach reinforce theoretical knowledge? 
  \begin{enumerate}
      \item[\textbf{RQ2.1}] What are the contents of the course that students perceive as reinforced by the integrated approach?
      \item[\textbf{RQ2.2}] What are the contents of the course in which we can perceive progression?
  \end{enumerate}
  \item[\textbf{RQ3}] How does the integrated approach allow students to experience the challenges of a real software development project?
  \begin{enumerate}
      \item[\textbf{RQ3.1}] What do students perceive as challenges faced?
      \item[\textbf{RQ3.2}] What do students perceive as challenges that are a good representation of a software development project?
  \end{enumerate}
\end{enumerate}

\subsection{Research Design}\label{researchdesign}

\begin{figure}[htb!]
\centering
\caption{Research path according to~\cite{wohlin2015towards}.}
\includegraphics[width=1\textwidth]{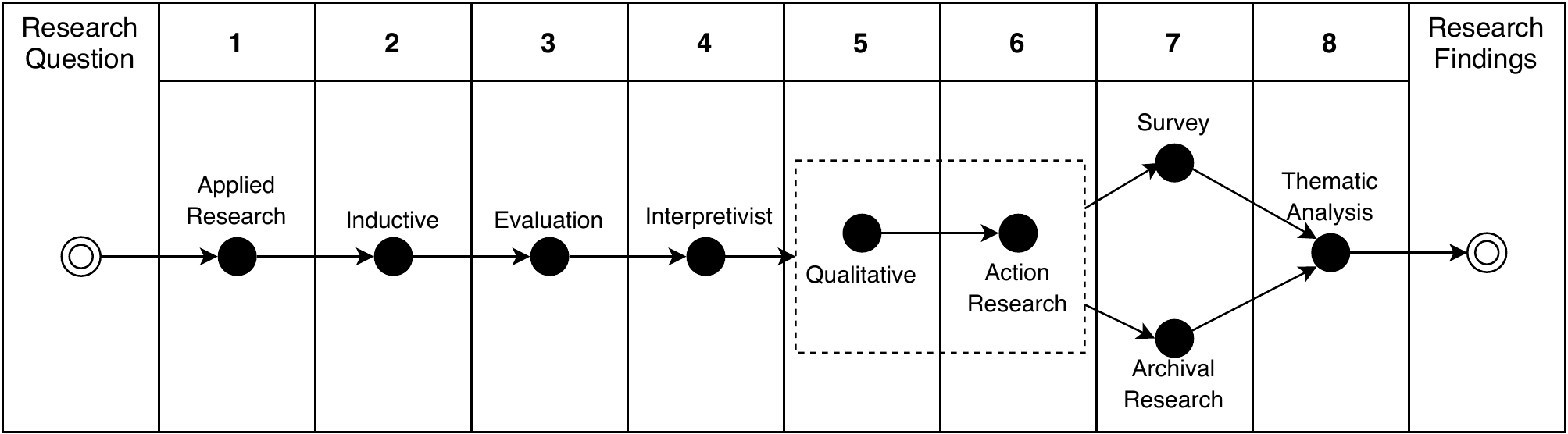}
\label{fig:researchpath}
\end{figure}

Our study is described according to the framework for selecting a research design in empirical software engineering by~\cite{wohlin2015towards}. We designed a research path (illustrated in Figure~\ref{fig:researchpath}) comprising a series of decision points that characterize our research.

On the strategic level (steps 1 to 4 in Figure~\ref{fig:researchpath}), our research is defined as (i) \textbf{applied} because it aims to provide solution to a specific problem; (ii) \textbf{inductive}, as we aim to identify patterns from observed data; (iii) with the purpose of \textbf{evaluate} the applicability of the integrated approach in the context; and it is (iv) \textbf{interpretivist} as it aims to understand the learning in the student's perspective.

The central unit of our work is qualitative action research (steps 5 and 6 in Figure~\ref{fig:researchpath}) that investigates a problem introduced by the pedagogical practice. We employed a set of data collection methods (step 7), including assessment rubrics, survey inquiry with students, and analysis of their written reports. The data was analyzed by open coding and thematic analysis (step 8).

\subsection{Data Collection}\label{datacollection}
We collected our data from the different TLAs and Assessment tasks planned in the course:
\begin{inparaenum}[1)]
\item from the successive submissions of the project plan written by the teams as part of the serious game; 
\item from the formative assessment of a survey created to gather students opinions about the integrated approach; as well as 
\item from the assessment of the reflection report teams wrote at the end of the course.
\end{inparaenum}
The multiple-source data collection approach aims at triangulating data from different sources and viewpoints. Through triangulation, we expected to identify different dimensions of the same phenomenon and to validate the data gathered from diverse sources. In the following subsections, we detail the different data sources we collected to assess the effectiveness of the approach to enable students learning.

\subsubsection{Assessment of Project Plan.}
\label{data_collection_1}
    
\begin{figure}[htb!]
\centering
\includegraphics[trim={2.3cm 3.3cm 2cm 3cm },clip,width=1\textwidth]{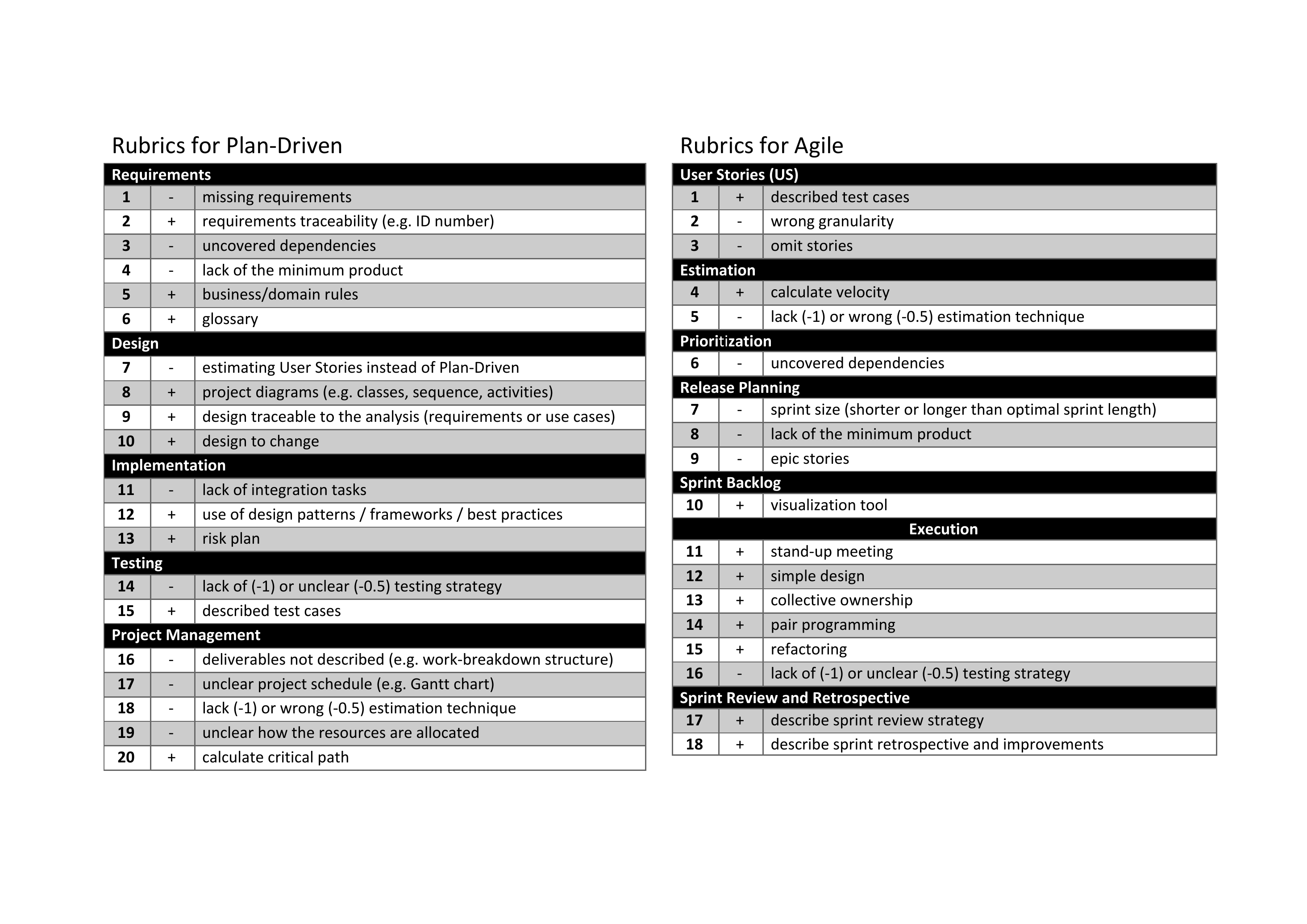}
\caption{Rubrics for  Plan-Driven (left) and Agile (right) game activities. The checklist items are based on good and bad practices for software projects covered by the course material.}
\label{fig:rubrics}
\end{figure}

The quantitative data was gathered during the game activity. Teams were requested to submit their project plans to be assessed before next workshop. Then, teachers reviewed the submissions and assessed them according to checklists (see rubrics for Plan-Driven and Agile projects in Figure~\ref{fig:rubrics}). The rubrics account for the number of good practices \textbf{(+)} and missing details \textbf{(-)} in the project plan. For each good practice demonstrated by the team, they scored +1 point, while for each bad practice, the team scored -1 point. A few checklist criteria also had partial scoring points, i.e., plus or minus 0.5 points, for partial achievement. 

The resulting scores from assessing the teams' projects were mainly used to calculate game modifiers the teams should apply for their dice rolls and condition the students' progress in the game. We informed the teams which checklist items they met and which ones they missed, so they have a guideline to improve their next submission. Therefore we expected teams to improve their project plans, and consequently, their performance after each round. The quantitative data collected from the project plans' assessment can help us answer research RQ2, and more particularly RQ2.2. 


\subsubsection{Formative Assessment.}
\label{data_collection_2}

After executing the integrated experiential learning approach, we conducted a retrospective meeting with the students to gather their opinion about the course and the approach. We created a survey questionnaire to collect data in a written format that we then used as data for assessing the effectiveness of the approach. The retrospective survey comprises ten open questions aligned to our research questions plus six questions aiming to improve the course, as detailed in Table~\ref{tab:survey}.

\begin{table}[htb!]
\centering
\scriptsize
\caption{Structure of the survey questionnaire and related research questions. The rightmost column maps the survey question to their related research questions.}
\label{tab:survey}
\begin{tabular}{lc}
\toprule
\textbf{Survey Questions} & \textbf{RQs} \\ \midrule
\textbf{About the Project Game} & \\
Q1. In your opinion, what is the objective of the games? & RQ1 \\
Q2. Which contents of the SE project course are reinforced by the games? & RQ2 \\
Q3. Which contents of the SE project course should be better represented by the games? & \\
Q4. Briefly describe the (three) most important difficulties / challenges you faced during the games. & RQ3 \\
Q5. In your opinion, what aspects (e.g. challenges) of a real project are represented in the games? & RQ3 \\ 
\midrule
\textbf{About the Development Task} & \\
Q6. In your opinion, what is the objective of the task? & RQ1 \\
Q7. Which contents of the SE project course are reinforced by the task? & RQ2\\
Q8. Which contents of the SE project course should be better represented by the task? & \\
Q9. Briefly describe the (three) most important difficulties / challenges you faced during the task. & RQ3\\
Q10. In your opinion, what aspects (e.g. challenges) of a real project are represented in the task? & RQ3 \\
\midrule
\textbf{Approaches Combined} & \\
Q11. In your opinion, what is the objective of combining the two approaches? & RQ1 \\
Q12. What have your learn with the approaches combined? & RQ2\\
\midrule
\textbf{Course evaluation} & \\
Q13. Please describe positive / negative aspects of the game. & \\
Q14. Please describe positive / negative aspects of the development environment. & \\
Q15. Please describe positive / negative aspects of the course schedule. & \\
Q16. What is your suggestion to improve the course activities? & \\ 
\bottomrule
\end{tabular}
\end{table}

The results from six survey questions (Q3, Q8, Q13-Q16) were merely used as feedback to the teaching staff to improve the course. The results from 2018's survey helped plan next year's application and, thus, impact how the integrated activity was conducted in 2019.

The remaining questions contributed to our research by providing the students' perception about (RQ1) the learning objectives of our integrated approach, (RQ2) theoretical knowledge reinforced by it, and (RQ3) challenges of a real software project.

\subsubsection{Written Reflection Report.}
\label{data_collection_3}

The last assessment task of the course was the submission of a reflection report. Students were asked to reflect on what was done during each step and keep notes about personal experiences during the integrated experiential learning approach. We encouraged the teams to critically discuss the causes of eventual deviations to their original plan within the groups.

By the end of the course, we invited students to compile the notes into a group reflection report and to write a post-mortem analysis of both teaching-learning approaches (game activity and development task) describing their experiences on the two different process models used, i.e., plan-driven and agile, plus their insights after the design-implement activity. The written report should also include a meta-reflection about the pros and cons of each development model. The open-text reflection contributed to answering our research questions RQ1, RQ2, and RQ3, similar to the formative assessment.

We provided the students with instructions for the written report, but we did not provide them a framework to follow. In that way, students were not limited by a formal structure and are fostered to express their thoughts more freely. The written reports were also used as an assessment task, and therefore part of the course grading. In our research, the assessment task was treated as an independent entity. The researchers conducting the data collection and data analysis did not participate in the grading of that particular assessment task.

\subsection{Data Analysis}

The three data sources provided us with both qualitative and quantitative data. We employed different methods to analyze each of them, as follows:

\subsubsection{Quantitative Data.} For the quantitative data, we assessed the teams' adherence to the good and the bad practices according to the checklists (see Section~\ref{data_collection_1}). We computed an overall score for each submission of a project plan, i.e., the difference between frequencies in which teams demonstrated good and bad practices. This calculation resulted in a score between -10 and 10; if the good practices were more frequent, the resulting score was positive, while more frequent bad practices resulted in negative scores.

By comparing the scores from two successive submissions of the project plans, we could assess the teams' progression and identify the criteria in which they progressed. This follow up allowed us to identify how many negative aspects were addressed, and how many positive aspects were reinforced with our game approach. The resulting data also helped teachers to identify theoretical content students did not understand well or could not express in their project plans. Such information is essential for our research, but it is also valuable for curriculum development and continuous improvement of the course.

\subsubsection{Qualitative Data.} The formative assessment and the written report provided mostly qualitative data we collected via coding analysis (see Sections~\ref{data_collection_2} and \ref{data_collection_3}). We employed thematic analysis for analyzing qualitative data, following the guidelines by \cite{Braun2006}. The thematic analysis comprises the coding and categorization of the textual information, data triangulation, and interpretation from different viewpoints~\citep{cruzes2011recommended}.

The coding process was carried out as follows: 

\begin{enumerate}
    \item One of the researchers (i.e., the second author) created the first iteration of the codebook by using theoretical concepts of software project management presented in the course lectures' materials. This codebook was organized as a hierarchical structure of themes, mapped to the course contents.
    \item The same researcher used this codebook for doing the first level coding. \citep{Saldana2015} of the survey responses, adding complementary codes that emerged from the survey responses. A list of the codes identified by the thematic analysis is provided in Appendix~\ref{appendixA}.
    \item Another researcher (i.e., third or fourth authors) used this updated codebook to do independent and blind coding of the same documents.
    \item After this, we aggregated codes from both researchers, and in the case where was a disagreement of coding in responses from the survey, the first author did the additional round of blind coding.
    \item We reached an agreement in cases in which two out of three researchers agreed on the same code. In three-way disagreement cases, all involved researchers participated in a joint discussion until reaching an agreement for the final code.
\end{enumerate}

Although we understand that qualitative thematic analysis's primary goal is not analyzing the frequency in which codes appear, we have computed code frequency for data synthesis. With this analysis, we can also assess how frequently two given codes co-occurred, i.e., occurred in the same piece of the text. By doing so, we identified the most frequent codes related to learning aspects (RQ1), course contents (RQ2), and students' experience (RQ3), thus helping us answering our these three research questions.

\subsection{Execution of the Data Collection and Analysis}

We conducted our empirical study alongside the two applications of the experiential approach, i.e., in 2018 and 2019. Each year, we collected data from the project plans, formative assessment, and written reports. The data collection instruments and data analysis follows the same procedure in both cases. Only the number of participants (students and teams) and one member of the teaching staff varied from the first year to the second.

\subsubsection{2018} Students enrolled in the course took part in our study as follows: 

\textbf{Assessment of project plan.} 14 teams participated in the game activity. During three rounds, they submit a project plan document. For the first and second instances, we assessed their project plans and provided feedback for improvement. Unfortunately, data for the second instance of the plan-driven approach is missing. Thus our analysis is limited to improvements during the agile project gameplay.

\textbf{Formative assessment.} We provided the students' explanation about each of the questions and the survey procedure. We further asked for their voluntary collaboration and ensured that no personal data is collected. The students took around 30 minutes to complete the survey.

45 out of 58 students were present and consented to our inquiry to respond to the electronic survey. Their answers provide textual feedback on their experience with the game, the development task, and the integration between the two approaches.

\textbf{Written reflection report.} All the 14 groups provided a post-mortem analysis of the integrated learning approach, compiling notes from two-game activities and the development task. Initially, two of the teams provided only a partial reflection (information about one of the exercises was missing). We contacted them and asked to complement the report, and they promptly did it.

\subsubsection{2019} Students enrolled in the course took part in our study as follows: 

\textbf{Assessment of project plan.} There were eight teams in the beginning of the course but one of the teams dispersed and their members joined other teams. Eventually, seven teams participated in the game activity. During four rounds (two for plan-driven and two for agile), they submit a project plan document. We provided assessment and feedback for their project plans in all the rounds, similar to last year.

\textbf{Formative assessment.} Similar to the previous year, we provided explanation for the students about each of the questions and the survey procedure. We further asked their voluntary collaboration and ensure that no personal data is collected. The students took around 30 minutes to complete the survey.

15 out of 24 students were present and consented with our inquiry to respond the electronic survey. Their answers provide textual feedback on their experience with the game, the development task, and the integration between the two approaches.

\textbf{Written reflection report.} All the seven groups provided a post-mortem analysis of the integrated learning approach, compiling notes from two game activities and the development task.

\section{Results}\label{sec:results}

This section presents the empirical study results in relation to our three data collection instruments (see Section \ref{datacollection}). A summary of the data collected for each instrument is available online, as complementary material~\citep{molleri2020dataset}.

\subsection{Assessment of Project Plan}
\begin{figure}[!ht]
\centering
\begin{subfigure}{.5\textwidth}
  \centering
  \includegraphics[width=1\textwidth]{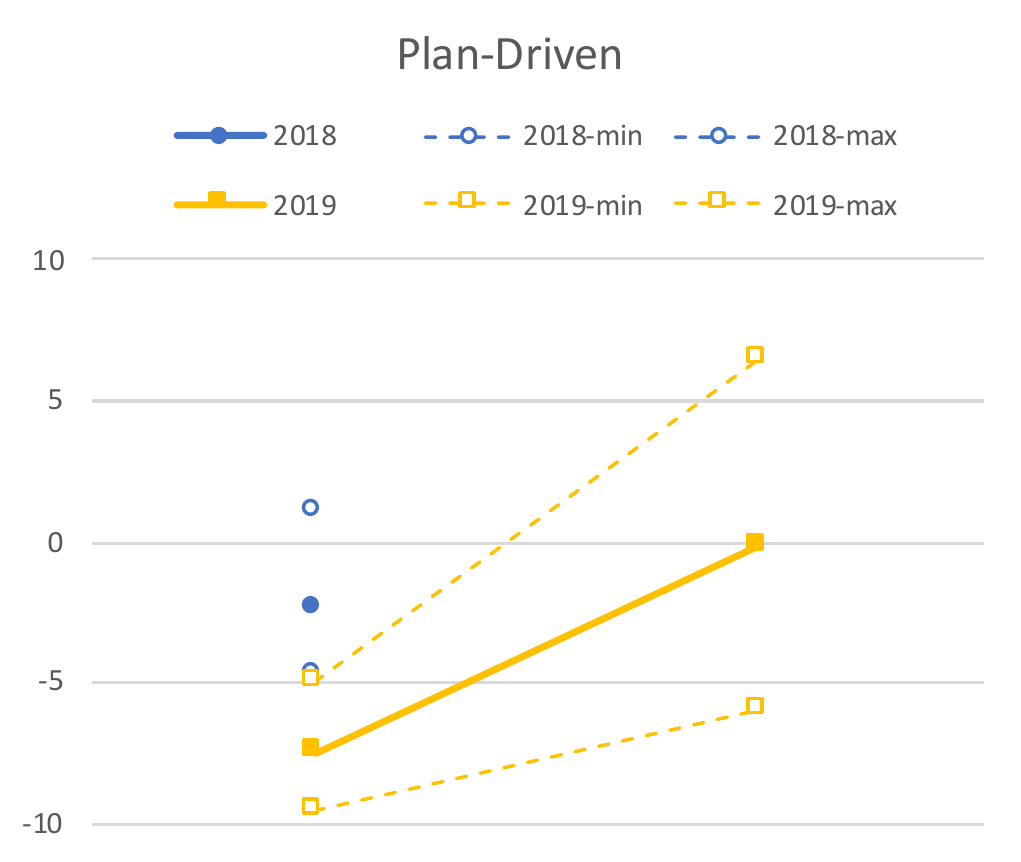}
  \caption{}
  \label{fig:balancePD}
\end{subfigure}%
\begin{subfigure}{.5\textwidth} 
  \centering
  \includegraphics[width=1\textwidth]{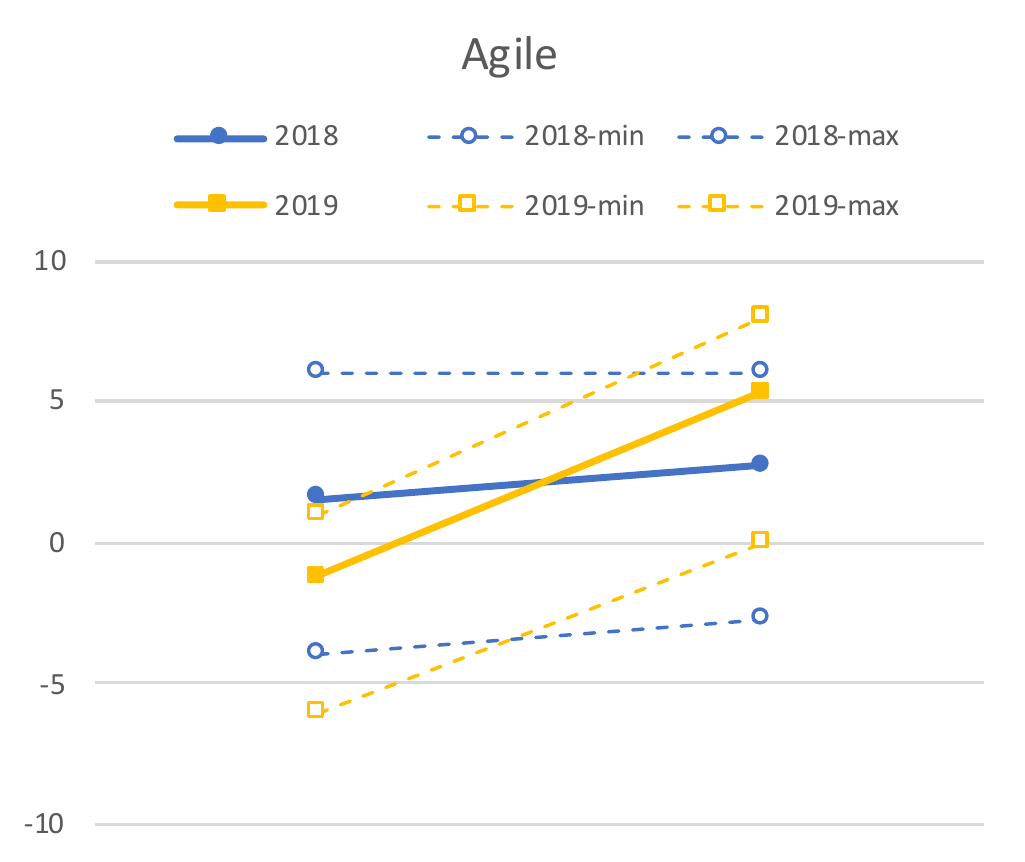}
  \caption{}
  \label{fig:balanceAgile}
\end{subfigure}
\caption{Average scores for first and second project submissions (left and right dots, respectively) in 2018 and 2019. A solid line shows the average improvement between the two assessments, while as the dotted lines above and below represent the maximum and minimum score obtained by a team.}
\label{fig:balanceBonuses}
\end{figure}

The results of our assessment of the teams' project plans using the checklist provided us with an indication of the students' progression about using good practices for software projects and avoiding the bad practices. Overall, in all instances of the game, there was a certain degree of improvement between the first and second submissions (see Figure~\ref{fig:balanceBonuses}). Each plot shows an aggregate balance of positive and negative scores for each instance of the experiential approach.

For the \textbf{Plan-Driven} (left plot), both 2018 and 2019 had a rough start; their average scores for the first submission were -2.41 and -7.5, respectively. That means that they included a high number of bad practices. It is worth to mention that the one team in 2018 performed positively in the first submission, with a score of +1. In the second submission, we could notice that all the teams in 2019 improved considerably, although the average score was still slightly negative, i.e., -0.125. Missing data from the Plan-Driven project in 2018 limit our analysis of the improvement only to 2019.

The teams started much better in the \textbf{Agile} (right plot), with an average score of 1.55 (2018) and -1.25 (2019). In 2018, the teams showed a slight improvement for the second submission, with an average score of 2.69. In 2019, the teams performed much better, raising their average scores to 5.31. None of the 2019 teams had an overall negative score for the second submission, the lowest one being 0. The results point out that a higher number of good practices were included in the Agile project plan than the Plan-Driven project plans.

A detailed presentation of the data for each criterion is provided in Table~\ref{tab:resultsRubrics}. Each row presents a criterion and points out the number of teams whose project plan fulfill it during the first and second assessment submissions. Each rubric criterion is graded with marks ranging from 0 or 1. In the majority of cases, 0 for absence or 1 for presence, except for the ones with a star (*) that could receive a partial value (0.5), e.g., RP14: lack of (-1) or unclear (-0.5) testing strategy.

\begin{table}[!ht]
\centering
\small
\caption{Results of the rubric assessment for Plan-Driven game (left) and Agile game (right) in 2018 and 2019. Each row represents a checklist criterion and if it relates to a good (+) or a bad (-) practice. (*) refers to criteria in which students can receive partial points values, i.e., $0.5$.}
\label{tab:resultsRubrics}
\begin{tabular}{lc|ccr}
\multicolumn{5}{c}{\textbf{Plan-Driven}} \\
\hline
 & & \multicolumn{3}{c}{\textbf{2019} (n = 8)} \\
\textbf{ID} & +/- & 1st & 2nd & $Prog$ \\  \hline
RP1 & - & 1 & 1 & 0\% \\ 
RP2 & + & 2 & 5 & +37.5\% \\ 
RP3 & - & 8 & 3 & +62.5\% \\ 
RP4 & - & 8 & 2 & +75\% \\ 
RP5 & + & 0 & 3 & +37.5\% \\ 
RP6 & + & 0 & 7 & +87.5\% \\ 
RP7 & - & 4 & 0 & +50\% \\ 
RP8 & + & 0 & 1 & +12.5\% \\ 
RP9 & + & 0 & 0 & 0\% \\ 
RP10 & + & 1 & 0 & \textcolor{gray}{-12.5\%} \\ 
RP11 & - & 8 & 6 & +25\% \\ 
RP12 & + & 0 & 1 & +12.5\% \\ 
RP13 & + & 0 & 6 & +75\% \\ 
RP14* & - & 7 & 1.5 & +34.4\% \\ 
RP15 & + & 0 & 3 & +37.5\% \\ 
RP16 & - & 8 & 6 & +25\% \\ 
RP17 & - & 7 & 2 & +62.5\% \\ 
RP18* & - & 6 & 3.5 & +15.6\% \\
RP19 & - & 7 & 5 & +25\% \\ 
RP20 & + & 1 & 3 & +25\% \\ 
\hline
\textbf{Average} & & & & \textbf{34.4\%} \\
\end{tabular}
\quad
\begin{tabular}{lc|ccr|ccr}
\multicolumn{8}{c}{\textbf{Agile}} \\
\hline
 & & \multicolumn{3}{c}{\textbf{2018} (n = 14)} & \multicolumn{3}{c}{\textbf{2019} (n = 7)} \\
\textbf{ID} & +/- & 1st & 2nd & $Prog$ & 1st & 2nd & $Prog$ \\  \hline
RA1 & + & 9 & 11 & +14.3\% & 1 & 3 & +28.6\% \\ 
RA2 & - & 3 & 4 & \textcolor{gray}{-7.1\%} & 2 & 0 & +28.6\% \\ 
RA3 & - & 8 & 6 & +14.3\% & 1 & 0 & +14.3\% \\ 
RA4 & + & 10 & 11 & +7.1\% & 1 & 3 & +28.6\% \\ 
RA5* & - & 0.5 & 0.5 & 0 & 3 & 0 & +28.6\% \\ 
RA6 & - & 3 & 3 & 0\% & 7 & 0 & +100\% \\ 
RA7 & - & 1 & 1 & 0\% & 2 & 0 & +28.6\% \\ 
RA8 & - & 2 & 4 & \textcolor{gray}{-14.3\%} & 6 & 0 & +85.7\% \\ 
RA9 & - & 3 & 2 & +7.1\% & 2 & 0 & +28.6\% \\
RA10 & + & 3 & 3 & 0\% & 4 & 4 & 0\% \\ 
RA11 & + & 5 & 6 & +7.1\% & 3 & 6 & +42.9\% \\ 
RA12 & + & 4 & 4 & 0\% & 3 & 4 & +14.3\% \\ 
RA13 & + & 1 & 1 & 0\% & 2 & 3 & +14.3\% \\ 
RA14 & + & 3 & 4 & +7.1\% & 1 & 5 & +57.1\% \\ 
RA15 & + & 0 & 2 & +14.3\% & 2 & 5 & +42.9\% \\ 
RA16* & - & 2 & 2 & 0\% & 4 & 0.5 & +25\% \\ 
RA17 & + & 3 & 7 & +28.6\% & 0 & 6 & +85.7\% \\ 
RA18 & + & 2 & 7 & +35.7\% & 0 & 4 & +57.1\% \\
\hline
\textbf{Average} & & & & \textbf{+6.3\%} & & & \textbf{39.1\%}\\
\end{tabular}
\end{table}

We also computed a progress variable $Prog$ to illustrate the teams' progress based on the feedback provided between both submissions. The progress was computed according to equation (\ref{eq:prog_equation}).

\begin{equation}
\label{eq:prog_equation}
Prog_{(criteria)} = (+/-) \frac{marks_{(2nd)} - marks_{(1st)}}{n}
\end{equation}

where:

\begin{itemize}
    \item $\textbf{+/-}$ corresponds to the positive or negative aspect of the given rubric criteria;
    \item $\textbf{marks}$ corresponds to the number of teams that demonstrated the criteria; and
    \item $\textbf{n}$ is total number of teams in the corresponding year .
\end{itemize}

As an example, in 2019, two teams included the good practice RP2: \textit{requirements traceability} in their first submission of the project plan, and five teams in the second submission; therefore of three out of eight teams improved their project plans, as shown in equation (\ref{eq:example1}). If we now focus on a negative criterion, in 2019, eight teams included the bad practice RP3: \textit{uncovery dependencies} in their project plans in their first submission. Only three teams included the bad practice in their second submission; therefore, five out of eight teams improved their project plans by removing the bad practice, as shown in equation (\ref{eq:example2}).

\begin{multicols}{2}                                               
    \noindent
    \begin{equation}
        Prog_{(RP2,2019)} = + \frac{5 - 2}{8} = +37.5\%
        \label{eq:example1}
    \end{equation}
    \begin{equation}
        Prog_{(RP3,2019)} = - \frac{3 - 8}{8} = +62.5\%
        \label{eq:example2}
    \end{equation}
\end{multicols}

Negative values in Table~\ref{tab:resultsRubrics} point out to a decrease in $Prog$ for good practices, i.e., positive criteria, meaning that instead of having more teams including the good practice in their second submission of the project plan, we have fewer teams including it. On the contrary, we can also have negative values pointing out a decrease on $Prog$ if fewer teams included a bad practice in their second submission of the project plan, we have more teams including it.

   \textbf{Plan Driven.} As shown in Table \ref{tab:resultsRubrics}, most aspects presented some degree of improvement. Just one aspect (RP10: \textit{design to change}) showed a slight decline: a group has included this practice in the first submission but removed such description in the next submission.

The highest improvement concern to RP6 \textit{using a glossary of terms}, RP4 \textit{detailing the minimum viable product}, and RP13 \textit{describing a risk plan}. It is important to notice that some good practices, e.g., RP4 and RP6, have not been included by any team in the first submission. After the students received feedback from the teaching staff, they improved their plans by considering such practices. None of the groups included RP9 \textit{design traceable to the analysis} in either submission. That points to a potential gap in how we introduce that topic during the lectures or in the game workshops.

\textbf{Agile.} We observed that the majority of the criteria presented some degree of improvement during the agile game. Exceptions are RA2 \textit{wrong granularity} and RA8 \textit{lack of minimum product}, that shown decline during the 2018 instance. Overall, improvements were greater in 2019, but the teams in 2018 had a better start, i.e., they covered more good practices and avoided the bad ones already in the first submission.

We observe a significant improvement in both years regarding criteria RA17 (describe sprint review strategy) and RA8 (detailing the minimum viable product). In 2019, the most substantial improvement concerns to RA6 (uncovered dependencies of tasks) with \textit{prog = 100\%}, i.e., all seven teams improved their project plans to address this criterion in their second submission.

There is only a criterion in which we observed no differences between the first and second submissions in both years, RA10: \textit{a visualization tool for sprint backlog}. In both years has the same values for the first and second submissions. The teams that employed this practice since the first submission kept using it, and the others did not adopt it in their second submission.

\subsection{Formative Assessment}

\begin{figure}[!ht]
\centering
\begin{subfigure}{.5\textwidth}
  \centering
  \includegraphics[trim={5cm 3cm 3cm 0},clip,width=1\textwidth]{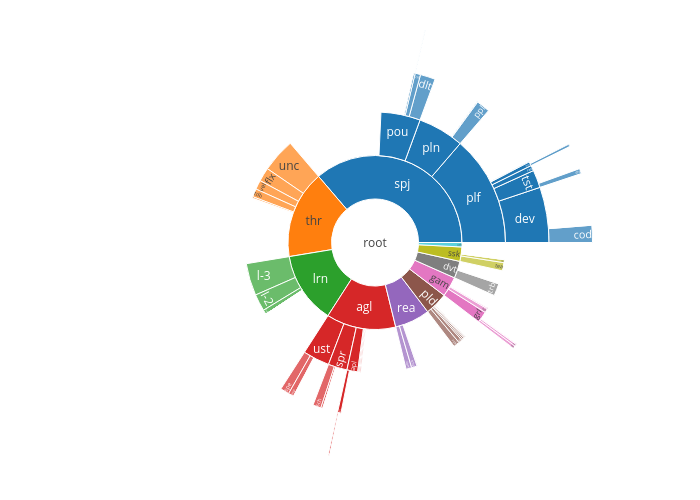}
  \caption{2018 (n = 45)}
  \label{fig:total2018}
\end{subfigure}%
\begin{subfigure}{.5\textwidth}
  \centering
  \includegraphics[trim={5cm 3cm 3cm 0},clip,width=1\textwidth]{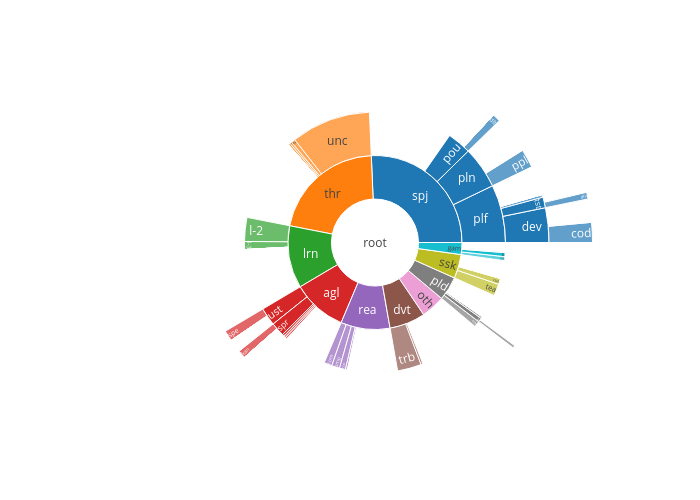}
  \caption{2019 (n = 15)}
  \label{fig:total2019}
\end{subfigure}
\caption{\textbf{Sunburst plots for relative frequency of codes in the formative assessment.} The size of each segment represents the number of times that code appear in relation to the total. Subcodes, a.k.a. leafs, also contribute to the size of their root code's segment, e.g. plf project lifecycle adds its frequency to spj software project.}
\label{fig:totalsunburst}
\end{figure}

For the formative assessment, we collected students' individual responses to survey questions (Table~\ref{tab:survey}) and coded them according to our code book (Appendix~\ref{appendixA}). Figures~\ref{fig:total2018} and \ref{fig:total2019} illustrate the frequency of codes identified in years 2018 and 2019, respectively.

For both years, spj sofware project and its leafs are the themes more frequently mentioned by the respondents. The relative size of the segments shows that it is more frequently mentioned in 2018 than in 2019. The second more frequently cited is \textit{thr theory}, which comprises flexibility, velocity, or uncertainty. In 2018, third and fourth places are \textit{agl agile}, which comprises understand, analyse, or experience and \textit{lrn learning}, respectively. In 2019, the order of these two was inverse.

Table~\ref{tab:surveyResults} summarize the top-3 more frequent codes related to each relevant survey question. In the bottom of the table, we summarize the top-5 codes from questions mapped to research questions RQ1 to RQ3. A full list of frequency values for each survey question is provided as complementary material \citep{molleri2020dataset}.

\begin{table}[!ht]
\centering
\small
\caption{Top-3 more frequent codes for each survey question and top-5 more frequent codes for the related research questions. Q3, Q8, Q13-Q16 are not presented in the table, as they are not intended to answer the research questions. Values in parenthesis show the frequency of the code with regards to that question.}
\label{tab:surveyResults}
\begin{tabular}{c|p{.4\textwidth}p{.4\textwidth}}
\toprule
 & \multicolumn{2}{c}{\textbf{Survey Questions}} \\
\midrule
\textbf{} & \textbf{2018 (n = 45)} & \textbf{2019 (n = 15)} \\ 
\midrule
Q1 & lrn learning (35), spj software project (32), l-2 understand (16) & lrn learning (8), thr theory (8), unc uncertainty (7) \\ 
Q2 & spj software project (18), agl agile (18), pdr plan driven (16) & thr theory (7), unc uncertainty (7), spj software project (3) \\
Q4 & agl agile (13), thr theory (12), spj software project (9) & spj software project (6), agl agile (5), pld plan driven (4) \\
Q5 & thr theory (12), spj software project (11), unc uncertainty (8) & thr theory (8), unc uncertainty (8), spj software project (3) \\
Q6 & spj software project (27), rea realism (24), lrn learning (23) & lrn learning (8), spj software project (8), other codes (3) \\ 
Q7 & spj software project (21), plf project lifecycle (15), agl agile (13) & agl agile (4), spj software project (3), other codes (2)  \\ 
Q9 & spj software project (17), dvt development task (13), trb troubleshooting (13) & dvt development task (6), trb troubleshooting (6), other codes (3) \\ 
Q10 & spj software project (12) plf project lifecyle (8), thr theory (8) & thr theory (6), unc uncertainty (6) spj software project (4) \\
Q11 & lrn learning (26), l-3 experience (20), other codes (16) & lrn learning (6), rea realism (5), spj software project (4) \\ 
Q12 & spj software project (16), thr theory (16), unc uncertainty (12) & rea realism (6), spj software project (5), com complexity (4) \\
\midrule
 & \multicolumn{2}{c}{\textbf{Research Questions}} \\
\midrule
\textbf{RQ1} & lrn learning (84), spj software project (70), rea realism (55), l-3 experience (51), thr theory (44) & lrn learning (22), spj software project (15), rea realism (14), thr theory (13), unc uncertainty (12) \\
\textbf{RQ2} & spj software project (50), thr theory (35), agl agile (31), plf project lifecycle (28), rea realism (23) & spj software project (10), thr theory (10), unc uncertainty (10), rea realism (7), agl agile (6) \\ 
\textbf{RQ3} & spj software project (49), thr theory (38), plf project lifecycle (28), agl agile (27), other codes (20) & thr theory (17), unc uncertainty (16), spj software project (14), agl agile (9), other codes (8) \\
\bottomrule
\end{tabular}
\end{table}

\subsubsection*{RQ1. Intended Learning Objectives}

Combining the results from both years, \textit{lrn learning}, \textit{spj software project}, and \textit{rea realism} are the codes more frequently related to RQ1. One student summarizes the objective of the learning approach as follows: \textit{``(To) get an experience on how projects work irl (in real life) and to get a feel for it''}.

Regarding the learning aspects, the game activity is more frequently related to \textit{l-2 understand}. In contrast, as \textit{l-3 experience} is more often related to the design-implement task (Q6), and to a combination of the two approaches (Q11). This shows an increasing level of complexity of the learning process, according to Bloom's taxonomy~\citep{krathwohl2009taxonomy}.


\subsubsection*{RQ2. Theoretical Knowledge Reinforced}

\begin{figure}[!ht]
\centering
\begin{subfigure}{.5\textwidth}
  \centering
  \includegraphics[trim={5cm 3cm 3cm 0},clip,width=1\textwidth]{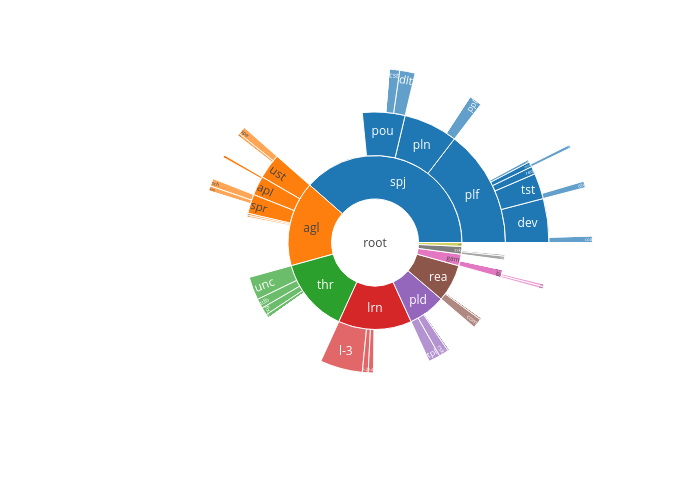}
  \caption{2018 (n = 45)}
  \label{fig:rq22018}
\end{subfigure}%
\begin{subfigure}{.5\textwidth}
  \centering
  \includegraphics[trim={5cm 3cm 3cm 0},clip,width=1\textwidth]{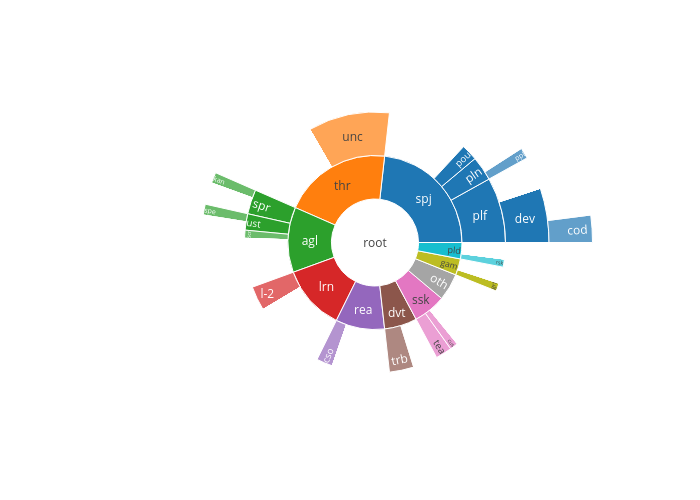}
  \caption{2019 (n = 15)}
  \label{fig:rq22019}
\end{subfigure}
\caption{Distribution of code frequencies for theoretical knowledge reinforced, i.e. sum of results of Q2, Q7, and Q12.}
\label{fig:rq2sunburst}
\end{figure}

A wide range of codes related to the theoretical content of the course have been mentioned by the participants. The sunburst plots~\ref{fig:rq22018} and \ref{fig:rq22019} cover most of the elements in our code book. Among them, \textit{spj software project}, \textit{agl agile} and \textit{thr theory} are among the five topmost for both years.

\begin{itemize}
    \item \textit{\textbf{spj software project:}} In our codebook, this code has the longest tree structure (see Appendix~\ref{appendixA}), so it is not surprising it was mentioned several times. The sub-codes that contributed most for this are \begin{inparaenum}[i)]
        \item \textit{pln project planning} and \textit{ppl proper planning};
        \item \textit{plf project lifecyle}, its sub-codes \textit{dev development}, and \textit{tst testing}; and
        \item \textit{pou project outcomes}, including \textit{cst costs} and \textit{dlt delivery time}.
    \end{inparaenum} Also, not surprisingly, project planning is more often mentioned in relation to the game activity (question Q2), and development in relation to the design-implement task (Q7). Overall, the student's perception pointed out aspects the integrated approach is designed to reinforce.
    \item \textit{\textbf{agl agile:}} \textit{ust user stories}, \textit{spr sprint} and \textit{apl agile planning} are the sub-codes more often mentioned. These three have been actively addressed by the second game instance and the design-implement task (Q2 and Q7).
    \item \textit{\textbf{pdr plan-driven:}} It is mostly mentioned in comparison to the agile method. For example, one student wrote \textit{``The things that are reinforced with the games are specially visualized with the agile model I would say. You've conducted the waterfall model in other projects, but the agile praxis was new and very educating.''}.
    \item \textit{\textbf{thr theory:}} Theories such as \textit{flx flexibility}, \textit{vel velocity}, and \textit{tdb technical debt} has been mentioned, but \textit{unc uncertainty} is overall three times more coded. Uncertainty has been perceived in all steps of our integrated approach. About the combination of the game and design-implement tasks, one student answered \textit{``the uncertainties in the game actually exist in real life projects''}.
\end{itemize}

\subsubsection*{RQ3. Challenges of a Real Project}

Out of the 5-topmost frequent codes, we noted a few similarities in both years: \textit{thr theory}, \textit{spj software project}, and \textit{agl agile}. We identify that different challenges are related to different steps of our approach, as follows: 

\begin{itemize}
    \item \textbf{Game activity:} \textit{spe story point estimation} and \textit{spl sprint planning} got particular attention during the game activity (reported by Q4). The students mentioned challenges such as \textit{``make small but relevant stories''}, \textit{``planning the sprint based on velocity''} and \textit{``prioritize what should be done''}. Challenges related to \textit{est estimation} are also mentioned for the plan-driven game, in particular as this was the first experience students had with such a task.
    \item \textbf{design-implement task:} \textit{cod coding} (a sub-code of \textit{spj software project}) was a particularly though challenge for the students during the design-implement task (Q9). In most cases, it is mentioned alongside \textit{trb troubleshooting}. For example, one student wrote \textit{``Challenges that were faced were things like trying to understand the task and how to implement and develop so it could go as smoothly as possible''}.
    \item \textbf{Integrated Experiential Learning Approach:} According to the students, \textit{unc uncertainty} (a sub-code of \textit{thr theory}) is the challenge better represented by both the game and design-implement tasks (Q5 and Q10). Some students wrote: \textit{``the uncertainty levels were especially well represented by the ever changing requirements''} and \textit{``in real life the project do not goes as planed''}.
\end{itemize}

\subsection{Reflection Report}

The written reflection report provides a comprehensive overview of all the aspects the students considered important during the course activity. The notes - took during the activity - describe their perception when they are carrying out the learning task, and their postmortem reflection shows insights and lessons learned. They also provide more details than the formative assessment, as the teams could discourse freely about their reasoning.

\begin{figure}[!ht]
\centering
\begin{subfigure}{.5\textwidth}
  \centering
  \includegraphics[trim={5cm 3cm 3cm 0},clip,width=1\textwidth]{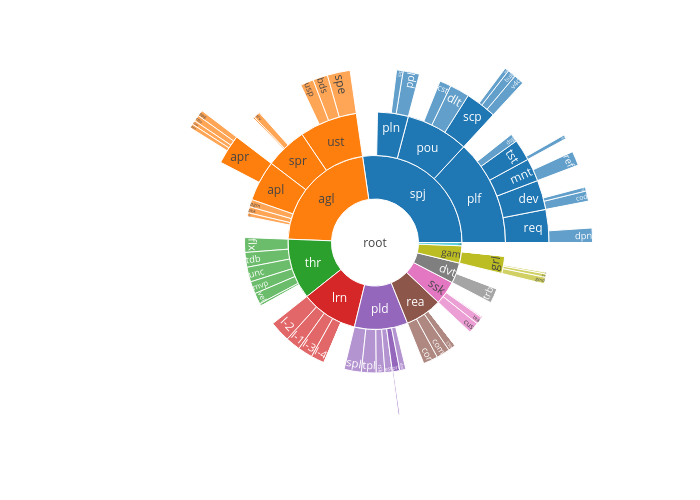}
  \caption{2018 (n = 14) }
  \label{fig:e2018}
\end{subfigure}%
\begin{subfigure}{.5\textwidth}
  \centering
  \includegraphics[trim={5cm 3cm 3cm 0},clip,width=1\textwidth]{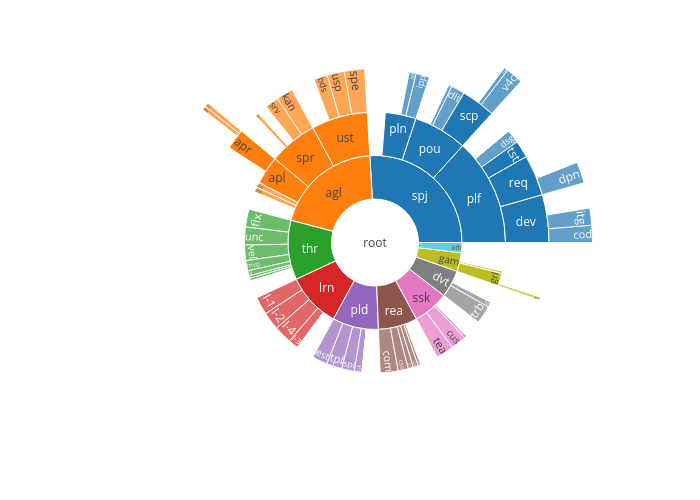}
  \caption{2019 (n = 7)}
  \label{fig:e2019}
\end{subfigure}
\caption{Summary of the codes on the reflection report across years. The size of the blocks correspond to the number of reports in which the code appears. We do not present here the frequency or extent in which the codes are discussed.}
\label{fig:totalsunburstreport}
\end{figure}

Using coding we could identify a wide range of aspects covered, as seen in Figure~\ref{fig:totalsunburstreport}. The results are consistent between 2018 and 2019 regarding the order and relative size of segments in the sunburst plots. Some codes were mentioned by all the teams, regardless of the year: \textit{spj software project}, \textit{plf project lifecycle}, \textit{agl agile}, \textit{ust user stories}, \textit{pld plan driven}, and \textit{rea realism}. Other codes were mentioned less often, as shown in the dataset (see complementary material \citep{molleri2020dataset}.

With regards to the different steps we instruct the students to reflect upon, results are as follows:

\begin{enumerate}[1)]
    \item \textbf{Plan-driven game:} The aspects more often mentioned by students are \textit{spj software project}, \textit{plf project lifecycle}, \textit{req requirements}, \textit{tst testing}, and \textit{sch schedule plan}. Much of the reflection is dedicated to requirements specification and to plan the workflow as a schedule. Project management tools such as Gantt charts were frequently reported. From a learning perspective, students relate their learning to \textit{l-1 remember} and \textit{l-2 understand}.
    \item \textbf{Agile game:} Cover a range of topics such as \textit{apl agile planning}, \textit{spr sprints}, \textit{ust user stories}, also \textit{spj software project}, \textit{plf project lifecycle}. It became clear the challenges students faced when estimating user stories and prioritize them for sprint planning. Theoretical aspects such as \textit{vel velocity}, \textit{mvp minimum viable product} and \textit{tdb technical debt} are also frequently discussed. On the pedagogical side, levels \textit{l-1 remember} and \textit{l-2 understand} are more often evidenced.
    \item \textbf{Comparing the Game Instances}: It is important to notice a few similarities between the two-game instances. Firstly, \textit{spj software project} and \textit{plf project lifecycle} are among the codes mentioned in all reports. Also, codes that reflect the corresponding aspects in both methodologies such as the pairs \textit{req requirements}/\textit{ust user stories} and \textit{sch schedule}/\textit{spr sprints}. Finally, the same learning levels, \textit{l-1 remember} and \textit{l-2 understand}, are identified in both cases, thus demonstrating the organization of ideas provided by the course's theoretical background.
    \item \textbf{Design-Implement task:} Codes that are mentioned more often include \textit{dvt development}, \textit{trb troubleshooting}, \textit{rea realism}, \textit{fle flexibility}, \textit{ust user stories}, and \textit{spj software project}. Three of these (i.e. ,\textit{rea realism}, \textit{ust user stories} and \textit{spj software project}) are also well mentioned in the agile game approach, thus pointing out the similarities between the two activities. From the theoretical aspects, \textit{unc uncertainty} is the most cited. On a pedagogical perspective, level \textit{l-3 apply} is more frequently mentioned, and some reports also showed \textit{l-4 analysis} elements. This points out an increasing complexity in the cognitive domain of learning.
    \item \textbf{Meta-reflection:} In this step, teams discussed their interpretation of the integrated learning approach and the main aspects they have learned. Both \textit{agl agile} and \textit{pld plan-driven} approaches are often discussed, as well as \textit{spj software project}, \textit{plf project lifecycle}. In addition to that, \textit{cso contextual solutions} has been extensively discussed, as students reflect which method best fit the scenarios they faced during the integrated approach. For example, one team summarizes \textit{``there is no one-way approach to solve a problem. Different projects would need different ways to be handled based on the demands of the project''}. Other theoretical aspects such as \textit{fle flexibility} and \textit{unc uncertainty} are also well cited. From the pedagogical stance, \textit{l-4 analysis} is the most covered, which is also a good signal since it shows a progression in the way concepts are aggregated and discussed.
\end{enumerate}

\section{Discussion}\label{sec:discussion}

The results provided in the previous section summarize the student's perception through three distinct data collection sources. In order to discuss the findings, we detailed them with regards to our three research questions:

\renewcommand{\thesubsection}{RQ\arabic{subsection}.}

\subsection{Intended Learning Objectives}

We can summarize students' perception regarding the ILOs of our approach by the three main codes associated with RQ1: \textbf{learning the real aspects of software projects}. The students perceived that the game approach is a simulation of the software process, and they could try different strategies to steer their projects. The simulation is grounded on the theoretical aspects they were instructed during the lectures. One student summarizes the objective as follows: \textit{``To simulate how our theoretical project would work out if applied to reality where there are uncertainties that can not be foreseen''}. The students' perception is, in our opinion, well-aligned to the course's intended learning objectives (see Section~\ref{sec:ilos}). Students also reported that the design-implement task provided them with real software development challenges, such as uncertainties and troubleshooting. The findings are supported by both the survey responses and the written report.

From Bloom's taxonomy perspective, the game activity is associated with understanding how the software process works. In contrast, the design-implement task is associated with applying or experiencing it in a practical context. Finally, the written report also provides evidence about higher cognitive tasks, such as determining relationships and making inferences based on their own experiences. These three levels of objectives from Bloom's taxonomy relates to codes l-2, l-3, and l-4. They are the same described by the course's ILOs.

In addition to this, students appreciated the experiential learning approach. We particularly noted their enthusiasm when rolling the dices for the game outcomes and when trying out their code with the LEGO Mindstorms\textsuperscript{TM} vehicles. Several students explicitly reported their appreciation in the formative survey and in the written report. Some even adopted roleplay to describe their experiences, such as a team discussing their product release: \textit{``Should we delay the delivery of our product or go ahead and launch our buggy product anyway? Since we don’t really know the weight and complexity of the bugs, it’s hard for us to evaluate how much they actually will affect the final product''}. 

\subsection{Theoretical Knowledge Reinforced}

Our findings show that students could recall and describe a wide range of theoretical knowledge. The sunburst diagrams (Figures~\ref{fig:rq2sunburst} and~\ref{fig:totalsunburstreport}) illustrate the comprehensiveness of the contents (RQ2.1) reported in the survey and written report, respectively.

The written report provides a more detailed description of reinforced knowledge, as students are asked to reflect and describe their experience. Although some aspects are better reinforced than others (e.g., agile is more frequently mentioned than plan-driven methodology), one can note that leaves from all main codes have been coded. That means that students not only mentioned the main themes but also discoursed different aspects of them. For example, a team provides the following reasoning for adopting the agile methodology: \textit{``Agile is requirement oriented, so it is flexible for the requirements changes that come during the processes''}. The team somehow successfully understood some of the theoretical characteristics of the methodology and related it to challenges (i.e., requirement changes) they experienced. One of the primary outcomes of the integrated approach is that, at the end of the design-implement task and in the reflection reports, the students express their awareness of uncertainty in software projects. In their reports, one can notice that they have ``connected the dots'': during the implementation task, they realize how the concept of uncertainty materializes in concrete events, like the troubleshooting they have to carry out during the design-implement tasks.

Regarding the assessment of teams' project plans, we used a set of pre-defined criteria to trace teams' progression (RQ2.2). Students have shown increasing compliance with good practices and reduced the presence of bad practices in their project plans between two assessment rounds. The evidence also points to improvements regarding the estimation and prioritization tasks. Our insight is that the game design helped the students develop a better understanding of such practices by experience, i.e., learning by doing, and therefore we can conclude that the approach itself supports students' progression.

\subsection{Challenges of Software Development}

The written report also provided a good understanding of students' challenges during the execution of our integrated approach (RQ3.2). Students reported that the rules and consequences of their decisions were not always clear to some of them during the game activity. Some students noted that this was the teaching staff's intention: to provide good examples of uncertainties they could not account for.

During the design-implement task, the main challenges reported by students are troubleshooting and unexpected technical issues. Interestingly, students were able to relate those issues to the uncertainties experienced during the game. The formative survey also supports those findings, as students reported that uncertainty and troubleshooting in the task are similar to real projects.

could relate challenges to other aspects of the software process. As an example, estimation and prioritization are challenges related to project planning. These two tasks showed significant improvement by assessing teams' project plans, pointing out that the integrated approach supports reinforcing knowledge through problem-solving. 


The experiential approach also helped students to realize the importance of proper planning. In a plan-driven project, it is essential to detail a project plan that accounts for potential deviations. In an agile project, one should be ready to adapt and change according to the new situations. An important finding relates to the contextual solutions, which is one of the central ``hidden'' learning outcomes of the course, i.e., no single solution will solve all challenges. Examples of students' reflections are \textit{``Depending on what type of project it is then the way of doing it is different''}, and \textit{``it is important to understand how to plan a software in the right context, taking into account the characteristics and limitations of the project''}.

\section{Threats to Validity}

\renewcommand{\thesubsection}{\arabic{section}.\arabic{subsection}}

Following the interpretivist paradigm, we describe threats to validity of our research according to the categories described by~\cite{lincoln2007naturalistic}:

\subsection{Credibility}

Our study aimed at evaluating an experiential learning approach from the perspective of the students. To achieve such a goal, we designed an empirical evaluation grounded on the Action Research method, i.e., our experiential learning approach acted as an intervention, as is described in~\ref{fig:researchpath}. We employed a mixed data collection approach to gathering a more holistic and rich picture of the phenomenon. To ensure that the participants could express their perceptions openly, we used an anonymous online questionnaire with open questions. There is a potential bias well-known to the participatory research, that the researcher could affect the data collection and analysis due its proximity to the phenomenon. We tried to mitigate such a threat by continuously engaging in joint discussions between the researchers acting as observers and others not directly involved in data collection. This multi-perspective reflection was crucial to challenge the view of the observers and make sense of the findings.

Another limitation relates to the number of instances in which we collected data for validating the approach. Although we ran the evaluation twice in successive years, the participants drew out of convenience, and thus we could not assure a wide variety of cultural characteristics and cognitive abilities. It is important also to mention that during the spring of 2020, we run the course again, with a bigger number of students from three different cohorts: five years Master of Science in Engineering program (300 ECTS) in Industrial Economy, five years Master of Science in Engineering program (300 ECTS) in AI and Machine Learning, and five years Master of Science in Engineering program (300 ECTS) in Software Development. However, while we were running the first instance of the serious game, we switched to distance learning due to the COVID19 outbreak. The change to distance learning imposed many challenges to both students and teaching staff.
On the one hand, we had to change several things on how we run the games (e.g., the workshops' structure, effective supervision time for each team), which might have added noise to the data gathered. On the other hand, and even more critical, the data gathering activities might represent additional work done by both the students and the teaching staff. Therefore, we decided to skip the data gathering and analysis in 2020. At the time of writing this paper, we are planning the course again in Spring 2021, but with much uncertainty regarding the extent to which we will meet students in serious game instances and the design-implement task.

\subsection{Confirmability}

The opinions of participants were coded in relation to the context of the course, according to a pre-conceived codebook. This codebook might have limited our initial analysis to the course's theoretical background, but we later expanded it using themes that emerged from participants' responses. We also acknowledge limitations in the researchers' ability to reflect upon the data due to experiences. However, we trust that researchers acting as teaching staff could interpret the students' opinions more appropriately. We also employed multiple coders, independent coding, and joint discussions to corroborated the findings among researchers.

Another limitation refers to our ability to present a chain-of-evidence from observation to findings. Due to concerns about anonymity, we are not able to make our complete data set available. Part of the qualitative data, in particular of the written report, could be traced to teams and/or individuals that participated in the course. Aiming to provide a transparent and accountable view of our data analysis process, we provide access to summaries of the assessment rubrics, and coding of the survey responses, and the reflection reports, as complementary material, available in~\citep{molleri2020dataset}
\footnote{The information are attached as appendixes to the manuscript for the double-blind review process, since the dataset contains traceable authorship information.}
.

\subsection{Transferability}
Our experiential approach tries to bridge the gap between theory and practice in the classroom. We acknowledge that some of the challenges we propose are artificial, fictional, yet still based on reality. The solutions in a professional environment might differ radically from the simulation, and students were aware of this constraint. One of the teams reflected on this limitation as follows: \textit{``we believe that there is a difference between professional planning and simulated planning in a scholarly environment. This was also the case for our project since we did not start to solve new problems until the ones started on was finished.''} In a professional environment, a sequential approach to solve problems is unrealistic as priorities change due to factors not covered in our approach, e.g., time to market, customer satisfaction, etc.

Although our validation results are meaningful in the context of the pedagogical application, they do not provide evidence supporting the use of this approach compared to other interventions, either other innovative learning approaches or the traditional lectures. We provide evidence of how the integrated experiential approach allows students to experience some of the challenges of software development projects and the aspects that the experimental approach reinforces. However, we can only make claims about the potential benefits of the approach in the studied context. Therefore we need to gather additional evidence to reinforce the results presented in this paper. We invite teachers on similar introductory courses to software engineering to replicate the setting with the serious games and the design-implement tasks and gather evidence of their interventions' results.

\section{Conclusions}\label{sec:conclusions}

This paper reported the execution and validation of an integrated approach to teaching SE courses using a problem-based, experiential learning approach. The approach integrates i) a serious game activity simulating the planning and execution of a software development project and ii) a design-implement task representing realistic, small-scale, software development project experience. The experiential approach builds on the software project's challenges simulated in the game and later represented in the design-implement task.

We executed the approach with students of a SE course in two consecutive years. To empirically validate the approach we coded and analyzed, by using thematic analysis, and frequency count: \begin{inparaenum} [i)]
    \item the adherence of their project plans to a set of rubrics that drive the serious game;
    \item the adherence of their project plans to a set of rubrics that drive the serious game; and
    \item the student’s perceptions collected through an online survey instrument
\end{inparaenum}

The results suggest that the experiential approach enables students' progression, is well aligned with the SE course's intended learning objectives, and it supports the acquisition of wide range of theoretical concepts and knowledge related to software projects and software engineering process models. 

The main contribution is provided by the integration between the game activity and the design-implement task. Although the students perceived them as distinct activities, they realized that some key contents of the course are reinforced by their integration, particularly the uncertainties of software development, the need for an appropriate plan, or how to deal with troublesome concepts such as relative estimation with story points. Besides, students acknowledge that the approach helped them infer the value of the contextual solutions in software projects.

We invite teachers on similar introductory courses to software engineering to replicate the setting with serious games (rules and rubrics can be found online), a controlled-scale, design-implement task, and evidence of the results of the intervention on the course.

\newpage
\appendix
\label{appendixA}
\section{Code book for Thematic Analysis}
\begin{figure}[hbt!]
\centering

\begin{tikzpicture}[mindmap, scale=0.5, transform shape,
  every node/.style={concept, execute at begin node=\hskip0pt},
  root concept/.append style={font=\small, minimum size=1.5cm},
  agile/.style={concept color=pink!40,faded/.style={concept color=pink!30}},
  project/.style={concept color=orange!40, faded/.style={concept color=orange!20}},
  learning/.style={concept color=purple!40, faded/.style={concept color=purple!20}},
  software_project/.style={concept color=green!40, faded/.style={concept color=green!20}},
  plan_driven/.style={concept color=violet!40, faded/.style={concept color=violet!20}},
  realism/.style={concept color=red!40, faded/.style={concept color=red!20}},
  theory/.style={concept color=blue!40, faded/.style={concept color=blue!20}},
  game/.style={concept color=cyan!40, faded/.style={concept color=cyan!20}},
  development/.style={concept color=green!40, faded/.style={concept color=green!20}},
  softskills/.style={concept color=yellow!40, faded/.style={concept color=yellow!20}},
  other/.style={concept color=orange!40, faded/.style={concept color=orange!20}},
  level 1 concept/.append style={font=\small, 
      level distance=130,
      sibling angle=40}
  ]
    
    \begin{scope}[mindmap, concept color=blue!40]
    \node [theory] (thr) at (-3,0) {\textbf{thr} Theory}[clockwise from=0] 
          child [theory, grow=210] {node (vel) {\textbf{vel} Velocity}}
          child [theory, grow=180] {node (tdb) {\textbf{tdb} Technical Debt}}
          child [theory, grow=150] {node (unc) {\textbf{unc} Uncertainty}}
          child [theory, grow=120] {node (fle) {\textbf{fle} Flexibility}}
    ;
    \end{scope}
    
    \begin{scope}[mindmap, concept color=purple!40]
    \node [learning] (lrn) at (12,-2) {\textbf{lrn} Learning}[clockwise from=0] 
          child [learning, grow=120] {node (l-a) {\textbf{l-a} Analyse}}
          child [learning, grow=90] {node (l-u) {\textbf{l-u} Understand}}
          child [learning, grow=60] {node (l-e) {\textbf{l-e} Experience}}
    ;
    \end{scope}
    
    \begin{scope}[mindmap, concept color=pink!40]
    \node [agile] (agl) at (-2,-12) {\textbf{agl} Agile}[clockwise from=0] 
          child [agile, grow=120] {node (pbk) {\textbf{pbk} Product Backlog}}
          child [agile, grow=150] {node (spr) {\textbf{spr} Sprints}
                child [agile, grow=125] {node (spl) {\textbf{spl} Sprint Planning}}
                child [agile, grow=165] {node (knb) {\textbf{knb} KANBAN}}
          }
          child [agile, grow=180] {node (pgm) {\textbf{pgm} Planning Game}}
          child [agile, grow=210] {node (ust) {\textbf{ust} User Stories}
                child [grow=210] {node (usp) {\textbf{usp} User Story Prioritization}}
                child [grow=165] {node (spe) {\textbf{spe} Story Points Estimation}}
          }
          child [agile, grow=240] {node (scr) {\textbf{scr} SCRUM}}
          child [agile, grow=270] {node (apl) {\textbf{apl} Agile Planning}}
    ;
    \end{scope}
    
    \begin{scope}[mindmap, concept color=orange!40]
    \node [project] (spj) at (4,-7) {\textbf{spj} Software Project}[clockwise from=0] 
          child [project, grow=150] {node (pou) {\textbf{pou} Project Outcomes}
                child [grow=135] {node (dlt) {\textbf{dlt} Delivery Time}}
                child [grow=180] {node (scp) {\textbf{scp} Scope}}
                child [grow=225] {node (cst) {\textbf{cst} Cost, Resources}}
          }
          child [project, grow=90] {node (plf) {\textbf{plf} Project Lifecycle}
                child [grow=180] {node (req) {\textbf{req} Requirements}}
                child [grow=135] {node (dsg) {\textbf{dsg} Design}}
                child [grow=90] {node (dev) {\textbf{dev} Development}
                    child [grow=90] {node (cod) {\textbf{cod} Coding}}
                }
                child [grow=45] {node (tst) {\textbf{tst} Testing / Verification}
                    child [grow=67] {node (dfx) {\textbf{dfx} Defect Fixing}}
                }
                child [grow=0] {node (mnt) {\textbf{mnt} Maintenance}
                    child [grow=45] {node (ref) {\textbf{ref} Refactoring}}
                }
          }
          child [project, grow=0] {node (pln) {\textbf{pln} Project Planning}
                child [grow=15] {node (ppl) {\textbf{ppl} Proper Planning}}
                child [grow=60] {node (ipp) {\textbf{ipp} Improve Project Plan}}
          }
    ;
    \end{scope}
    
    \begin{scope}[mindmap, concept  color=violet!40]
    \node [plan_driven] (pdr) at (14,-13) {\textbf{pdr} Plan Driven}[clockwise from=0] 
          child [plan_driven, grow=150] {node (tpl) {\textbf{tpl} Traditional Planning}}
          child [plan_driven, grow=180] {node (ran) {\textbf{ran} Risk Analysis}}
          child [plan_driven, grow=210] {node (sch) {\textbf{sch} Schedule Plan}}
          child [plan_driven, grow=240] {node (doc) {\textbf{doc} Documentation}}
    ;
    \end{scope}
    
    \begin{scope}[mindmap, concept color=red!40]
    \node [realism] (rea) at (-4,-20) {\textbf{rea} Realism}[clockwise from=0] 
          child [realism, grow=180] {node (sim) {\textbf{sim} Simplicity}}
          child [realism, grow=210] {node (cso) {\textbf{cso} Contextual Solutions}}
          child [realism, grow=240] {node (com) {\textbf{com} Complexity}}
    ;
    \end{scope}
    
    \begin{scope}[mindmap, concept color=yellow!40]
    \node [softskills] (ssk) at (3,-16) {\textbf{ssk} Soft skills}[clockwise from=0] 
          child [softskills, grow=0] {node (tea) {\textbf{tea} Teamwork}}
          child [softskills, grow=30] {node (cus) {\textbf{cus} Customer}}
    ;
    \end{scope}
    
    \begin{scope}[mindmap, concept color=cyan!40]
    \node [game] (gam) at (1,-20) {\textbf{gam} Game}[clockwise from=0] 
          child [game, grow=300] {node (grl) {\textbf{grl} Game Rules}
                child [grow=270] {node (gtr) {\textbf{gtr} Game Turns}}
                child [grow=315] {node (gmd) {\textbf{gmd} Game Modifier}}
                child [grow=360] {node (gop) {\textbf{gop} Game Optimization}}
          }
          child [game, grow=270] {node (fun) {\textbf{fun} Fun, Appreciation}}
          child [game, grow=240] {node (fai) {\textbf{fai} Fairness}}
    ;
    \end{scope}
    
    \begin{scope}[mindmap, concept color=green!40]
    \node [development] (dvt) at (6,-20) {\textbf{dvt} Development Task}[clockwise from=0] 
          child [development, grow=330] {node (fun) {\textbf{fun} Fun, Appreciation}}
          child [development, grow=360] {node (trb) {\textbf{trb} Troubleshooting}}
    ;
    \end{scope}
    
    
    \begin{pgfonlayer}{background}
    \path (vel) to[circle connection bar switch color=from (blue!40) to (pink!40)] (spr);
    \path (pln) to[circle connection bar switch color=from (orange!40) to (pink!40)] (apl);
    \path (pln) to[circle connection bar switch color=from (orange!40) to (violet!40)] (tpl);
    \path (dlt) to[circle connection bar switch color=from (orange!40) to (pink!40)] (spl);
    \path (cst) to[circle connection bar switch color=from (orange!40) to (violet!40)] (ran);
    \path (scp) to[circle connection bar switch color=from (orange!40) to (pink!40)] (pbk);
    \path (dvt) to[circle connection bar switch color=from (green!40) to (cyan!40)] (gam);
    \end{pgfonlayer}

\end{tikzpicture}

\caption{Map of the codes representing the elements of our codebook. Links from the same color connect codes to their subcodes. Links that change color describe relation between elements from different code trees.}
\label{fig_map}

\end{figure}
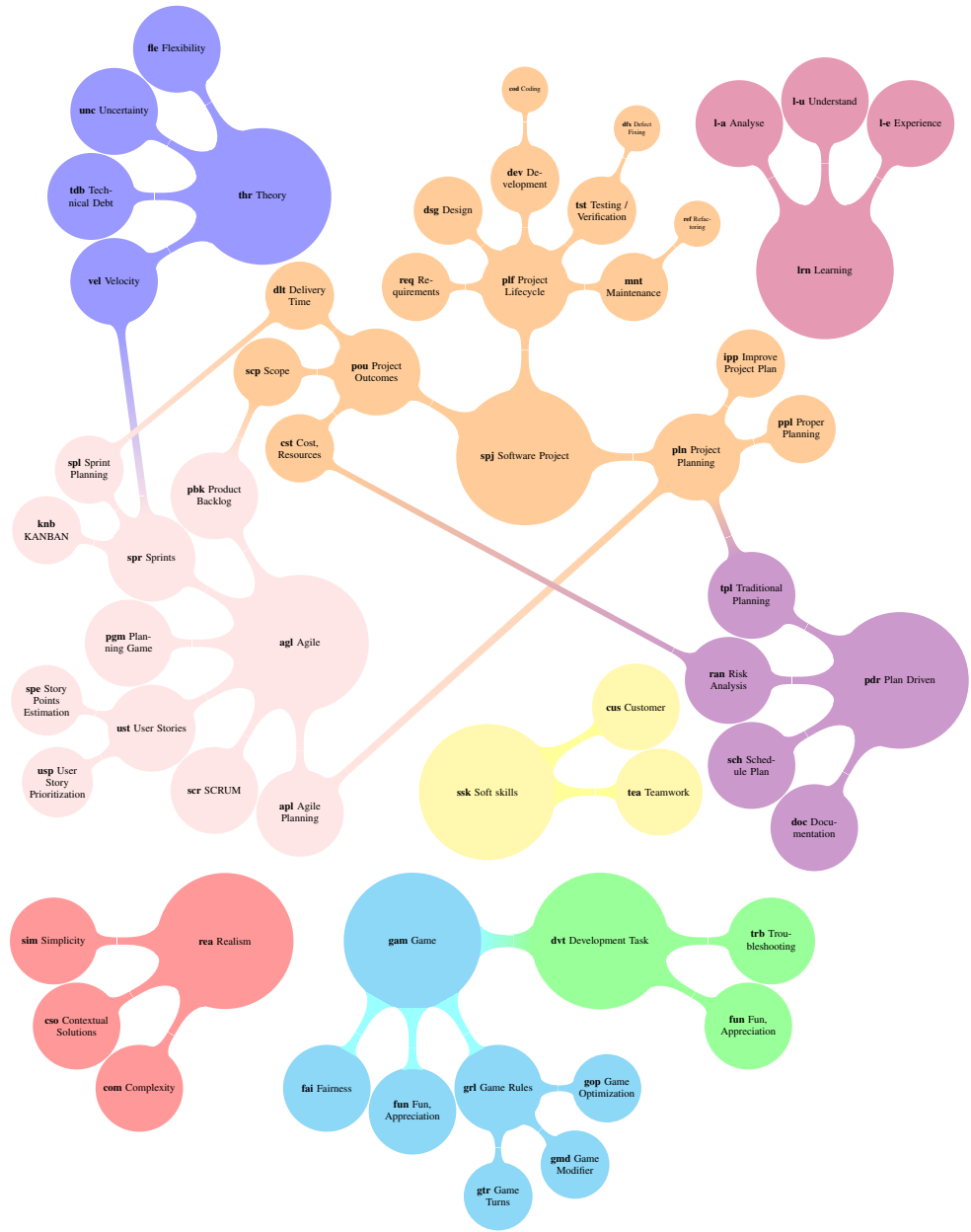


\newpage

\bibliographystyle{spbasic} 
\bibliography{references}

\end{document}